\pdfoutput=1 
\documentclass[12pt]{article}
\usepackage[numbers]{natbib}
\usepackage{a4}
\usepackage{amsmath}
\usepackage{amssymb}
\usepackage{latexsym}
\usepackage{amssymb}
\usepackage{epsfig}
\usepackage{natbib}
\usepackage{bm}
\usepackage{color}
\def \ii{{\mathrm{i}}}

\def \d{{\mathrm{d}}}

\def \pd{\partial}

\def \e{{\mathrm{e}}}

\def \BR{{\boldsymbol{R}}}

\def \BJ{{\boldsymbol{J}}}

\def \BB{{\boldsymbol{B}}}
\def \BE{{\boldsymbol{E}}}
\def \BH{{\boldsymbol{H}}}
\def \BD{{\boldsymbol{D}}}
\def \BA{{\boldsymbol{A}}}

\def\onedot{$\mathsurround0pt\ldotp$}
\def\cdddot#1{
  \mathbin{\vcenter{\baselineskip.67ex
    \hbox{\onedot}\hbox{\onedot}\hbox{\onedot}%
  }}%
}

\begin{document}
\title{{\bf 
Second gradient electromagnetostatics:\\ 
electric point charge, electrostatic and magnetostatic dipoles}}
\author{
Markus Lazar$^\text{}$\footnote{
Corresponding author. 
{\it E-mail address:} lazar@fkp.tu-darmstadt.de.
} 
\ and Jakob Leck 
\\ \\
        Department of Physics,\\
        Darmstadt University of Technology,\\
        Hochschulstr. 6,\\      
        D-64289 Darmstadt, Germany\\
}

\date{\today}    
\maketitle

\begin{abstract}
In this paper,  we study the theory of second gradient electromagnetostatics as the static version 
of second gradient electrodynamics.  
The theory of second gradient
electrodynamics is a linear generalization of higher order of
classical Maxwell electrodynamics  whose Lagrangian
is both Lorentz and $U(1)$-gauge invariant.
Second gradient electromagnetostatics is a  gradient field theory with up to 
second-order derivatives of the electromagnetic field strengths in the Lagrangian.
Moreover, it possesses a weak nonlocality in space and gives a regularization based on higher-order partial differential equations. 
From the group theoretical point of view, 
in second gradient electromagnetostatics
the (isotropic) constitutive relations involve an invariant scalar differential operator of fourth order in addition to scalar constitutive parameters.
We investigate the classical static problems of an electric point charge, and electric and magnetic dipoles in the framework of second gradient electromagnetostatics,
and we show that all the electromagnetic fields (potential, field strength, interaction energy, interaction force)
are singularity-free unlike the corresponding solutions in the classical Maxwell electromagnetism as well as in the Bopp-Podolsky theory. 
The theory of second gradient electromagnetostatics  delivers a singularity-free electromagnetic field theory with weak spatial nonlocality. 
\\ 

\end{abstract}

\section{Introduction}

In recent years, there has been a continuous interest in the so-called Bopp-Podolsky theory \citep{Zayats,Perlick2015,Perlick19,Bonin2019,Ji,AS19,Lazar19}, a first-order linear gradient theory of electrodynamics. 
The theory was introduced in the early 1940s by Bopp and Podolsky \citep{Bopp,Podolsky,PS} as a method to remove singularities in classical Maxwell electrodynamics and to obtain a consistent Lorentz- and gauge-invariant theory of point charges with finite self-energy (see also \citep{Lande}), thereby proposing an alternative to earlier theories that achieved the same goal through a nonlinear generalization of electrodynamics \citep{BI1934}.
While the motivation was ultimately the quantization of the theory (e.g. \citep{Podolsky2}), it is, first of all, a classical field theory. Through the introduction of an additional gradient term in the Lagrangian the generalized Maxwell equations yield linear partial differential equations of fourth order for the electromagnetic potentials.
Along with the additional term a new constant has to be introduced, the Bopp-Podolsky length scale parameter $\ell$, which by the original idea was supposed to be related to the electron self-energy. Indeed, \citet{Iwan,Kvasnica} and \citet{Cuzi} reasoned that the Bopp-Podolsky length scale parameter $\ell$ is of the order of $\sim 10^{-15}$\,m which is the order of the classical electron radius, however, \citet{AM97}, and \citet{Perlick19} argued that the Bopp-Podolsky length scale parameter $\ell$ should be equal to or smaller than $\sim 10^{-18}$\,m.
From the mathematical point of view the length scale parameter $\ell$ plays the role of a regularization parameter.
The regularization through the higher-order field equations provides a finite self-energy for the point charge, and its electrostatic potential is finite and nonsingular, its electric field finite, but with a directional discontinuity. Also, in Bopp-Podolsky electrodynamics the electromagnetic fields of a non-uniformly moving point charge possess a directional discontinuity on the light cone.
In order to obtain electromagnetic fields of a non-uniformly moving point charge 
with no directional discontinuity on the light cone, 
the theory of second gradient electrodynamics
has recently been proposed and used in~\citep{Lazar20}.
Further important 
advantages of gradient electrodynamics in comparison to the classical Maxwell electrodynamics are:  
no infamous ``4/3-problem", no unphysical runaway solutions to the equation of motion of a moving charged particle (the analogue of the Lorentz-Dirac equation), 
no singularities of the electromagnetic fields on the light cone, the self-force of a non-uniformly moving charged particle is regular (see also the discussion in~\citep{Lazar20}).

Besides the point charge, other important textbook examples of the Maxwell theory are the electrostatic
and magnetostatic dipoles possessing typical dipole singularities 
($1/r^3$- and Dirac delta-singularity)
in the electromagnetic fields~\cite{Jackson,Frahm,Griffiths,Leung}.  
In the mathematical literature, there is an interest in the 
regularization of the dipole singularities arising from the second-order
derivatives of $1/r$ in the sense of generalized functions 
(see, e.g., \citep{Hnizdo,Estrada13,Muniz}).  
The ideal magnetostatic dipole in first-order gradient electrodynamics was already studied by Land\'e and Thomas \cite{Lande3}, giving the magnetic fields and a finite self-energy. The latter result, however, turns out to be erroneous.
Using the Bopp-Podolsky theory, 
the electric and magnetic fields of electrostatic and magnetostatic dipoles
are still singular and their self-energy is also infinite, as will be
shown in this paper.  

Our purpose  is to investigate the theory of second gradient electromagnetostatics which is the static version of 
the theory of second gradient electrodynamics~\citep{Lazar20}.
We will study the textbook examples of electric point charge, electrostatic dipole and magnetostatic dipole in the  framework of generalized electrodynamics, and show that second gradient electromagnetostatics yields nonsingular dipole fields and gives a straightforward regularization of the 
dipole singularities based on higher-order partial differential equations.

In general, in generalized electrodynamics,
the electromagnetic fields (electric and magnetic potential, electric and magnetic field strengths) should satisfy  the following conditions:
\begin{itemize}
\item
the field must be finite at $r=0$,
\item
the field must be everywhere continuous,
\item
the self-energy of the field must be finite. 
\end{itemize}
As mentioned, not all conditions can be satisfied for an electric point charge, an electrostatic dipole and a magnetostatic dipole
using the Bopp-Podolsky electrodynamics.

Nowadays,
generalized continuum theories and in particular gradient continuum theories are very popular in physics, applied mathematics, material science
and engineering science. 
Gradient continuum theories are continuum theories which might possess characteristic length scales and characteristic time scales 
in order to describe size effects and memory effects, respectively.  
In particular, gradient theories are continuum theories valid at small scales unlike classical continuum theories like  Maxwell electrodynamics.
Because classical continuum theories are not valid at small scales, they lead to unphysical singularities at such scales. 
Thus, we are forced to regularize at short distances the classical continuum theories by means of generalized continuum theories.  
Gradient continuum theories provide nonsingular solutions of the field equations and
a regularization of classical singularities is achieved.
In physics, the most popular gradient continuum theory is the Bopp-Podolsky theory ~\citep{Bopp,Podolsky},
which is the first-order gradient version of the theory of electrodynamics as mentioned above.
In engineering science, a very popular gradient continuum theory is Mindlin's theory of first strain gradient elasticity~\citep{Mindlin64} . 
An advantage of gradient elasticity theory is that it can be connected with atomistic theories and all material parameters including the appearing length scale parameters
can be determined from ab initio calculations and using atomistic potentials (see, e.g.,  \citep{Shodja,Po,PAL}). 
Exciting gradient effects, which are important for applications in material science, exist due to the coupling between gradient elasticity and gradient electricity
in gradient electroelasticity~\citep{KA}, 
like flexoelectricity in solids \citep{flexo} which is the property of a dielectric material whereby it exhibits a spontaneous electrical polarization induced by an elastic strain gradient.
Furthermore, \citet{Mindlin65} introduced the 
theory of second strain gradient elasticity 
(see also~\cite{Jaunzemis,LMA06,AL09}). 
Mindlin's theory of second strain gradient elasticity 
involves additional material constants,
in addition to the elastic constants, which can be determined from atomistic potentials (see \citep{Shodja2}). 
Second strain gradient elasticity provides a better modelling of atomistics than first strain gradient elasticity.
Using a simplified version of second gradient elasticity, it was possible to obtain nonsingular solutions for the elastic field produced by point defects
which are elastic dipoles in solids \citep{L19}.
The theory of second gradient electrodynamics has been recently proposed by~\citet{Lazar20}.  
It turns out that second gradient electrodynamics provides a better mathematical modelling of electromagnetic fields at small distances than 
the Bopp-Podolsky electrodynamics (first gradient electrodynamics).
In this paper, we study the static version of it
called second gradient electromagnetostatics. 
Of course, the coupling between second strain gradient elasticity  and second gradient electromagnetostatics
may lead to many interesting gradient effects of higher order which will be worth to study more in detail in future work. 
Therefore, gradient continuum theories are very exciting research areas of physics on small scales. 

While structurally, as mathematical theories, gradient electrodynamics and gradient elasticity of $n$-th order are analogous, their physical significance differs slightly. Both can serve the purpose of regularization at small scales, but while gradient elasticity can be interpreted to describe microstructure and can, for example, also be derived as an approximation of lattice theories (see, e.g., \citep{Mindlin65,ASS02}), 
in gradient electrodynamics analogous interpretations are not as clear. Moreover, the length scale for gradient elasticity is of the order of $10^{-10}\,\text{m}$ and thus, as mentioned above, can be compared with atomistic simulations, however, in gradient electrodynamics the smallness of the length scale parameter has so far eluded experimental verification. While possible approaches have been suggested (e.g. \citep{Cuzi}), so far none has reached the scale of $10^{-15}\,\text{m}$ or smaller for the Bopp-Podolsky parameter and comparisons with quantum mechanical effects have yielded upper estimates for the length scale parameter (e.g. \citep{AM97, Perlick19}). 
However, gradient electrodynamics remains an interesting subject, as candidate for a consistent classical field theory of electrodynamics including point charges (see also \citep{Kiess2019}), as candidate for a generalized quantum electrodynamics (see, e.g., \citep{Bufalo11,Bufalo12}) 
and in comparison with other mathematical techniques of regularization.

The outline of this paper is as follows.
In Section~\ref{sec2}, the theory of 
second gradient electromagnetostatics is presented. 
In Section~\ref{sec3}, we give the collection of all relevant 
Green functions and their derivatives.
In Section~\ref{sec4}, the nonsingular electromagnetic fields of 
a point charge, an electrostatic dipole and a magnetostatic dipole
are computed in the framework of second gradient electromagnetostatics.
The limit of those electromagnetic fields to the Bopp-Podolsky 
theory and to the classical Maxwell theory are given in Section~\ref{sec5}
and Section~\ref{sec6}, respectively. 
The conclusions are given in Section~\ref{concl}.

\section{Second gradient electromagnetostatics}
\label{sec2}
In this Section, we provide the theoretical framework 
of second gradient electromagnetostatics.
Second gradient electromagnetostatics is the static version of 
second gradient electrodynamics\footnote{For details of second gradient electrodynamics we refer to~\citep{Lazar20}.}
given in~\citep{Lazar20}.
In the theory of second gradient  electromagnetostatics, 
the electrostatic and magnetostatic fields are described by the Lagrangian density
\begin{align}
\label{L-BP}
{\cal L}_{\text{grad}}&=
\frac{\varepsilon_0}{2}\, 
\Big(\bm E \cdot \bm E 
+\ell_1^2 \nabla \bm E :\nabla \bm E 
+\ell_2^4 \nabla\nabla \bm E \mathbin{\vdots} \nabla\nabla \bm E 
\Big)\nonumber\\
&\ 
-\frac{1}{2\mu_0 }\, 
\Big(\bm B \cdot \bm B 
+\ell_1^2 \nabla \bm B :\nabla \bm B 
+\ell_2^4 \nabla\nabla \bm B \mathbin{\vdots} \nabla\nabla \bm B 
\Big)
-\rho\phi+\bm J\cdot \bm A\,,
\end{align}
with the notation 
$ \nabla\nabla \bm E \mathbin{\vdots} \nabla\nabla \bm E 
=\pd_k\pd_j E_i \pd_k \pd_j E_i$,
$ \nabla \bm E :\nabla \bm E =\pd_j E_i \pd_j E_i$ and
$ \bm E \cdot\bm E =E_i E_i$. 
Here 
$\phi$ is the electrostatic scalar potential, $\BA$ is the magnetostatic vector potential,
$\BE$ is the electrostatic field strength vector,
$\BB$ is the magnetostatic field strength vector,
$\rho$ is the electric charge density, and 
$\BJ$ is the electric current density vector. 
$\varepsilon_0$ is the electric constant 
and 
$\mu_0$ is the magnetic constant 
(also called permittivity of vacuum
and permeability of vacuum, respectively).
Moreover, $\ell_1$ and $\ell_2$ are 
the two (positive) characteristic length scale parameters in second gradient 
electrodynamics 
and $\nabla$ is the vector operator Del (or Nabla).
In addition to the classical terms, first and second spatial derivatives of the (static) electromagnetic field strengths   ($\BE$, $\BB$) 
multiplied by the characteristic lengths $\ell_1$ and $\ell_2$ appear
in Eq.~\eqref{L-BP} describing a weak nonlocality in space.

While in classical electrodynamics and electromagnetostatics the requirements of isotropy and gauge invariance lead to a unique choice for the Lagrangian, we here only have uniqueness up to null-Lagrangians (cf.~\citep{Olver}). Bopp \citep{Bopp} and Podolsky \citep{Podolsky} introduced first-order gradient electrodynamics using different Lagrangians both being equal up to null-Lagrangians and leading to identical field equations. Our choice of Lagrangian is closer to Bopp's convention, using contractions of field gradients rather than divergences. Of course, with the introduction of higher-order terms the number of possible null-Lagrangians increases.

Note that, as is the case in classical electromagnetostatics with linear constitutive relations, the Lagrangian~\eqref{L-BP} is a sum of two purely electrostatic or magnetostatic terms. In consequence, unlike in Born-Infeld electromagnetostatics \citep{Kiess2011}, it is obvious that electrostatics and magnetostatics are separated: electric currents do not produce electric fields and electric charges do not produce magnetic fields. Also note that the two energy densities are positive definite, which results in positive definite energy functionals and thus in well-posed variational problems. While in the second gradient term the positive sign is both necessary and sufficient for positivity of the energy functional, in the first term it is only sufficient. As long as $\ell_1^4 < 4 \ell_2^4$ a negative sign could be allowed 
from the mathematical point of view, however, as will be seen below (case (3)), this would be unphysical.
While the static theory works formally with this choice of parameters, the dynamic generalization contains serious problems. 
Also, the Lagrangian with a negative sign in the first gradient term would not be a generalization of the Bopp-Podolsky theory where the positive sign is mandatory.

In electromagnetostatics, 
the electromagnetic field strengths ($\BE$, $\BB$)
can be expressed in terms
of the static electromagnetic  potentials ($\phi$, $\BA$)
\begin{align} 
\label{E}
\BE&=-\nabla \phi\,,\\
\label{B}
\BB&=\nabla\times \bm A\,
\end{align}
because they satisfy the two electromagnetostatic Bianchi identities
\begin{align}
\label{BI-1}
\nabla\times\BE&=0\,,\\
\label{BI-2}
\nabla\cdot \BB&=0\,,
\end{align}
which are known as the homogeneous Maxwell equations.
Eq.~\eqref{BI-1} states that the electrostatic field $\bm E$ is irrotational,
and Eq.~\eqref{BI-2} states that the magnetostatic field $\bm B$ has no scalar sources.

The Euler-Lagrange equations of the Lagrangian~\eqref{L-BP} with respect to the scalar potential $\phi$ and the 
vector potential $\bm A$ give the electromagnetic field equations
\begin{align}
\label{EL-1}
& L(\Delta)\,
\nabla\cdot \bm E=\frac{1}{\varepsilon_0}\,\rho\,,\\
\label{EL-2}
&
 L(\Delta)\, \nabla\times\bm B=\mu_0\,\BJ\,,
\end{align}
respectively, and 
the scalar differential operator of fourth order is given by
\begin{align}
\label{L-op}
 L(\Delta)=1-\ell_1^2 \Delta +\ell_2^4 \Delta^2\,,
\end{align}
where $\Delta$ is the Laplacian.
Eqs.~\eqref{EL-1} and \eqref{EL-2} are the generalized inhomogeneous
Maxwell equations in second gradient electromagnetostatics 
which are partial differential
equations of fifth order.
Eq.~\eqref{EL-1} represents the generalized Gauss law, 
and Eq.~\eqref{EL-2} represents the generalized Amp{\`e}re law.
The electric current density vector 
fulfills the equation of continuity
\begin{align}
\label{CE}
&\nabla\cdot \BJ =0\,.
\end{align}

If we use the variational derivative with respect to 
the electromagnetic fields ($\bm E$, $\bm B$),
then we obtain the (isotropic) constitutive relations in second gradient electromagnetostatics 
for the response quantities
($\bm D$, $\bm H$) in vacuum 
\begin{align}
\label{CE1}
\BD&:=\frac{\delta{\cal L_{\text {grad}}}}{\delta \BE}=\varepsilon_0\, 
 L(\Delta)\,
\BE\,,\\
\label{CE2}
\BH&:=-\frac{\delta{\cal L_{\text {grad}}}}{\delta \BB}=\frac{1}{\mu_0}\,
 L(\Delta)\,
\BB\,,
\end{align}
where $\BD$ is the electric excitation vector and  
$\BH$ is the magnetic excitation vector.
Therefore, in second gradient electromagnetostatics
the (isotropic) constitutive relations~\eqref{CE1} and \eqref{CE2} involve an invariant scalar constitutive operator of fourth order, $L(\Delta)$,
in addition to the scalar constitutive parameters $\varepsilon_0$ and $\frac{1}{\mu_0}$.
The constitutive operator $L(\Delta)$ is the only linear scalar isotropic operator of fourth order, 
a fact that is related to the uniqueness up to null-Lagrangians of the Lagrangian for the theory. 
Constitutive operators of this form already showed up in second strain gradient elasticity (e.g.~\citep{Mindlin65,LMA06,L19}).
The higher-order terms in Eqs.~\eqref{CE1} and \eqref{CE2} describe the polarization
of the vacuum present in second gradient electrodynamics (see, e.g., \citep{Santos2011,Lazar20}). 
Using the constitutive relations~\eqref{CE1} and \eqref{CE2}, 
the Euler-Lagrange equations~\eqref{EL-1} and \eqref{EL-2} 
can be rewritten in the form of inhomogeneous Maxwell equations
\begin{align}
\label{ME-inh}
\nabla\cdot \BD&=\rho\,,\\
\label{ME-inh2}
 \nabla\times\BH&=\BJ\,. 
\end{align}

From Eqs.~(\ref{EL-1}) and (\ref{EL-2}),
the following inhomogeneous partial differential equations, being partial differential
equations of sixth order, can be derived for the static electromagnetic field strengths
\begin{align}
\label{E-w}
 L(\Delta)\,
\Delta\,\BE&=\frac{1}{\varepsilon_0}\, \nabla\rho\,,\\
\label{B-w}
 L(\Delta)\,
\Delta\,\BB&=-\mu_0\,\nabla\times\BJ\,.
\end{align}

Using the generalized Coulomb gauge condition\footnote{Here the standard Coulomb gauge yields the same results, the necessity for the generalized condition only arises in quantum field theories corresponding to the theory presented here.} (see \citep{GP,Lazar2014,Bonin2019})
\begin{align}
\label{LG}
L(\Delta)\, \nabla \cdot \bm A=0\,,
\end{align}
the electromagnetic gauge potentials fulfill the following 
inhomogeneous partial differential equations of sixth order
\begin{align}
\label{phi-w}
 L(\Delta)\,
\Delta\,\phi&=-\frac{1}{\varepsilon_0}\, \rho\,,\\
\label{A-w}
L(\Delta)\,\Delta\,\bm A&=- \mu_0\, \BJ\,.
\end{align}

The differential operator of fourth order~\eqref{L-op}  
can be written in the form as  product of two Helmholtz operators with two length scale parameters $a_1$ and $a_2$,
which is called bi-Helmholtz operator,
\begin{align}
\label{L-op-2}
L(\Delta)=\big(1-a_1^2\Delta\big)\big(1-a_2^2\Delta\big)
\end{align}
with
\begin{align}
\label{a1a2-1}
\ell_1^{2}&=a_1^{2}+a_2^{2}\, ,\\
\label{a1a2-2}
\ell_2^{4}&=a_1^{2}\, a_2^{2}\,
\end{align}
and 
\begin{align}
\label{a1-2}
a^{2}_{1,2}&=\frac{\ell_1^{2}}{2}\Bigg(1\pm\sqrt{1-4\,\frac{\ell_2^{4}}{\ell_1^{4}}}\Bigg)\,.
\end{align}

The two length scales $a_1$ and $a_2$ may be real or complex.
In the theory of second gradient electromagnetostatics,  
the condition for the character, real or complex, of the two lengths 
$a_1$ and $a_2$ 
is the condition for the discriminant in Eq.~\eqref{a1-2}, $1-4\ell_2^4/\ell_1^4$, 
to be positive or negative.
Depending on the character of the two length scales $a_1$ and $a_2$
one can distinguish between the following cases:
\begin{itemize}
\item[(1)]  
$\ell_1^4>4\ell_2^4$\,:\\
The length scales 
 $a_1$ and $a_2$ are real and distinct and they read 
\begin{align}
\label{a1-2-2}
a_{1,2}&=\ell_1\,\sqrt{\frac{1}{2}\pm \frac{1}{2}\,\sqrt{1-4\left(\frac{\ell_2}{\ell_1}\right)^{\!4}}}
\end{align}
with $a_1>a_2$.
The limit to the Bopp-Podolsky theory  is given by  $\ell_2^4\rightarrow 0$. 
\item[(2)]
$\ell_1^4=4\ell_2^4$\,: \\
The length scales $a_1$ and $a_2$ are real and equal
\begin{align}
 a_1 =a_2=\frac{\ell_1}{\sqrt{2}}=\ell_2\,.
 \end{align}  
There is no limit to the Bopp-Podolsky theory.
This case can lead to Green functions having a time dependence that increases or decreases slowly, which can give rise to unphysical results
(e.g. \citep{Pais,Lazar20}).
\item[(3)]
$\ell_1^4<4\ell_2^4$\,:\\
The two length scales $a_1$ and $a_2$  are complex conjugate
\begin{align}
\label{c1-2-c}
a_{1,2}&=
A\pm\ii B\,,
\end{align}
 with
\begin{align}
\label{AB}
A=\ell_2\,\sqrt{\frac{1}{2}+\frac{\ell_1^2}{4\ell_2^2}}\,,\qquad
B=\ell_2\,\sqrt{\frac{1}{2}-\frac{\ell_1^2}{4\ell_2^2}}\,.
\end{align}
There is no limit to the Bopp-Podolsky theory.
For generalized electrodynamics,
this case leads to Green functions having a time dependence that increases exponentially, an 
acausal propagation and complex mass terms (e.g.~\citep{Pais,MS,Lazar20}). 
The dispersion relations of the vacuum, analogous to those computed in \citep{Santos2011}, have complex coefficients, suggesting instabilities or dissipation in the vacuum.\\
The possible negative sign in the first gradient term of the Lagrangian mentioned above also yields complex $a_1$ and $a_2$ and thus has similar consequences.
\end{itemize}
Therefore, the case~(1) is the physical one
and is the generalization of the Bopp-Podolsky theory (first gradient electromagnetostatics)
towards second gradient electromagnetostatics.

\section{Green functions in second gradient electromagnetostatics}
\label{sec3}

Second gradient electromagnetostatics is a linear theory 
with partial differential equations of sixth order, and
the method of Green functions (fundamental solutions) 
can be used.

The Green function $G^{L\Delta}$ of the sixth order differential operator 
$L(\Delta)\, \Delta$ is defined by 
\begin{align}
\label{BPE}
 L(\Delta)\,
\Delta\, G^{L\Delta}(\bm R)=\delta(\bm R)\,,
\end{align}
where $\bm R= \bm r-\bm r'$ and $\delta$ is the Dirac delta-function. 
The partial differential equation of sixth order~(\ref{BPE}) 
can be written as an equivalent  system of partial differential equations of
lower order 
\begin{align}
\label{BPE-2}
 L(\Delta)\,
G^{L\Delta}(\bm R)&=G^\Delta(\bm R)\, ,\\
\label{wave}
\Delta\, G^\Delta(\bm R)&=\delta(\bm R) \,,
\end{align}
or alternatively
\begin{align}
\label{BPE-3}
\Delta\, G^{L\Delta}(\bm R)&=G^{L}(\bm R)\, ,\\
\label{KGE}
 L(\Delta)\,
G^{L}(\bm R)&=\delta(\bm R)\,,
\end{align}
where $G^\Delta$ is the Green function of the Laplace operator~\eqref{wave} and 
$G^{L}$ is the Green function of the bi-Helmholtz operator~\eqref{KGE}.

 Using partial fraction decomposition, 
 the inverse differential operators 
 $\big[L(\Delta)\big]^{-1}$ and
 $\big[L(\Delta) \Delta\big]^{-1}$  with Eq.~\eqref{L-op-2}
 read in the formal operator notation (see also \citep{Schwartz})
\begin{align}
\label{Deco-L}
\big[L(\Delta)\big]^{-1}=\frac{1}{a_1^2-a_2^2}
\Big(a_1^2\, \big[1-a_1^2 \Delta\big]^{-1}
-a_2^2\, \big[1-a_2^2 \Delta\big]^{-1}\Big)
\end{align}
and
\begin{align}
\label{Deco-L2}
\big[L(\Delta)\Delta\big]^{-1}=\Delta^{-1}
+\frac{1}{a_1^2-a_2^2}
\Big(a_1^4\, \big[1-a_1^2 \Delta\big]^{-1}
-a_2^4\, \big[1-a_2^2 \Delta\big]^{-1}\Big)\,.
\end{align}
This formal notation directly translates into relations for the Green functions, so that
the Green function $G^{L}$ can be
written as a linear combination of two Green functions $G^{\rm H}(a_1)$ and  $G^{\rm H}(a_2)$
corresponding to the two length scale parameters $a_1$ and $a_2$ and Helmholtz operators
$[1-a_1^2\Delta]$ and $[1-a_2^2\Delta]$
\begin{align}
\label{KGG}
G^{L}=\frac{1}{a_1^2-a_2^2}
\Big(a_1^2\, G^{{\rm H}}(a_1)-a_2^2\, G^{\rm H}(a_2)\Big)\,.
\end{align}
Similarly, the Green function $G^{L\Delta}$ can be written 
as a linear combination of the Green function $G^\Delta$ of the
Laplace operator 
and  the two Green functions $G^{\rm H}(a_1)$ and  $G^{\rm H}(a_2)$
corresponding to the two length scale parameters $a_1$ and $a_2$ 
\begin{align}
\label{BPG}
G^{L\Delta}=G^\Delta
+\frac{1}{a_1^2-a_2^2}
\Big(a_1^4\, G^{{\rm H}}(a_1)-a_2^4\, G^{\rm H}(a_2)\Big)\,.
\end{align}
Note that the foregoing is essentially an application of proposition 1.4.4 in \citep{OW2015} and could analogously be applied in linear gradient theories of any order.
Using Eq.~\eqref{KGG}, the Green function of the bi-Helmholtz equation might be derived from 
the Green function of the Helmholtz equation\footnote{Sometimes, the differential operator, $L=1-\ell^2 \Delta$, is called modified Helmholtz operator~\citep{Zauderer}
or metaharmonic operator~\citep{Ortner}.}  (see, e.g., \citep{Kanwal,Zauderer,Schwartz}).
Therefore, the bi-Helmholtz field is a superposition of two Helmholtz fields with length scales $a_1$ and $a_2$. 
Using Eq.~\eqref{BPG}, the Green function of the bi-Helmholtz-Laplace equation might be derived by using the expressions of 
the Green function of the Laplace operator (see, e.g., \citep{Barton,Wl,Schwartz}) and
the Green function of the Helmholtz operator  (see, e.g., \citep{Kanwal,Zauderer,Schwartz}).
Therefore, the  bi-Helmholtz-Laplace field is a superposition of the Laplace field
and two Helmholtz fields. 

On the other hand, resulting from the decomposition into the systems Eqs.~\eqref{BPE-2} and \eqref{wave}, or \eqref{BPE-3} and \eqref{KGE}, 
the Green function of the bi-Helmholtz-Laplace operator can be written as the
convolution of the Green function of the Laplace operator and 
the Green function of the bi-Helmholtz equation
\begin{align}
\label{GBP-conv}
G^{L\Delta}=G^\Delta*G^{L}\,.
\end{align}
Here, the symbol $*$ denotes the spatial convolution.
Therefore, the Green function $G^L$ plays the role of the regularization function in second gradient electromagnetostatics. 
Moreover, the Green function of the bi-Helmholtz equation can be written as
convolution of the Green functions of the two Helmholtz operators 
\begin{align}
\label{GKG-conv}
G^{L}=
G^{{\rm H}}(a_1)*G^{\rm H}(a_2)\,,
\end{align}
satisfying Eqs.~\eqref{KGE} and \eqref{L-op-2}.

\subsection{Green functions}

The (three-dimensional) Green functions (or fundamental solutions) of the 
Laplace operator~\eqref{wave}, the Helmholtz operator with length parameter $a_1$, 
the bi-Helmholtz operator~\eqref{KGE} and the bi-Helmholtz-Laplace operator 
are given by
\begin{align}
\label{G-w-3d}
G^\Delta(\bm R)&=-\frac{1}{4\pi R}\,,\\
\label{G-KG-3d}
G^{\rm H}(\bm R)
&=\frac{1}{4\pi a_1^2 R}\,\text{e}^{-R/a_1}\,,\\
\label{G-BKG-3d}
G^{L}(\bm R)
&=\frac{1}{4\pi(a_1^2-a_2^2) R}\,
\Big(\e^{-R/a_1}-\e^{-R/a_2}\Big)\,,\\
\label{G-BP-3d}
G^{L\Delta}(\bm R)
&=-\frac{1}{4\pi R}\,
\bigg(1-\frac{1}{a_1^2-a_2^2}\,
\Big[a_1^2 \e^{-R/a_1}-a_2^2 \e^{-R/a_2}\Big]\bigg)\,.
\end{align}
Eq.~\eqref{G-BKG-3d} is obtained by substituting  Eq.~\eqref{G-KG-3d} into Eq.~\eqref{KGG},
and 
Eq.~\eqref{G-BP-3d} is obtained by substituting  Eqs.~\eqref{G-w-3d} and \eqref{G-KG-3d} into Eq.~\eqref{BPG}.
Moreover, the Green function~\eqref{G-BP-3d} may be written as
\begin{align}
\label{G-BP-2}
G^{L\Delta}(\bm R)
&=-\frac{1}{4\pi R}\, f_0(R,a_1,a_2)
\end{align}
with the auxiliary  function
\begin{align}
\label{f0}
f_0(R,a_1,a_2)=1-\frac{1}{a_1^2-a_2^2}\,
\Big[a_1^2 \e^{-R/a_1}-a_2^2 \e^{-R/a_2}\Big]\,.
\end{align}
The series expansion (near field) 
of the auxiliary function~\eqref{f0} reads as 
\begin{align}
\label{f0-ser}
f_0(R,a_1,a_2)&=\frac{1}{(a_1+a_2)}\, R-\frac{1}{6 a_1a_2(a_1+a_2)}\, R^3+\mathcal{O}(R^4)\,.
\end{align}
Therefore, the function $f_0(R,a_1,a_2)$ 
regularizes up to a $1/R$-singularity towards a nonsingular field expression. 
Indeed, the Green function~\eqref{G-BP-3d} is nonsingular and finite at $R=0$, namely 
\begin{align}
\label{G-BB-0}
G^{L\Delta}(0)=-\frac{1}{4\pi (a_1+a_2)}\,.
\end{align}

\begin{figure}[t]\unitlength1cm
\vspace*{0.1cm}
\centerline{
\epsfig{figure=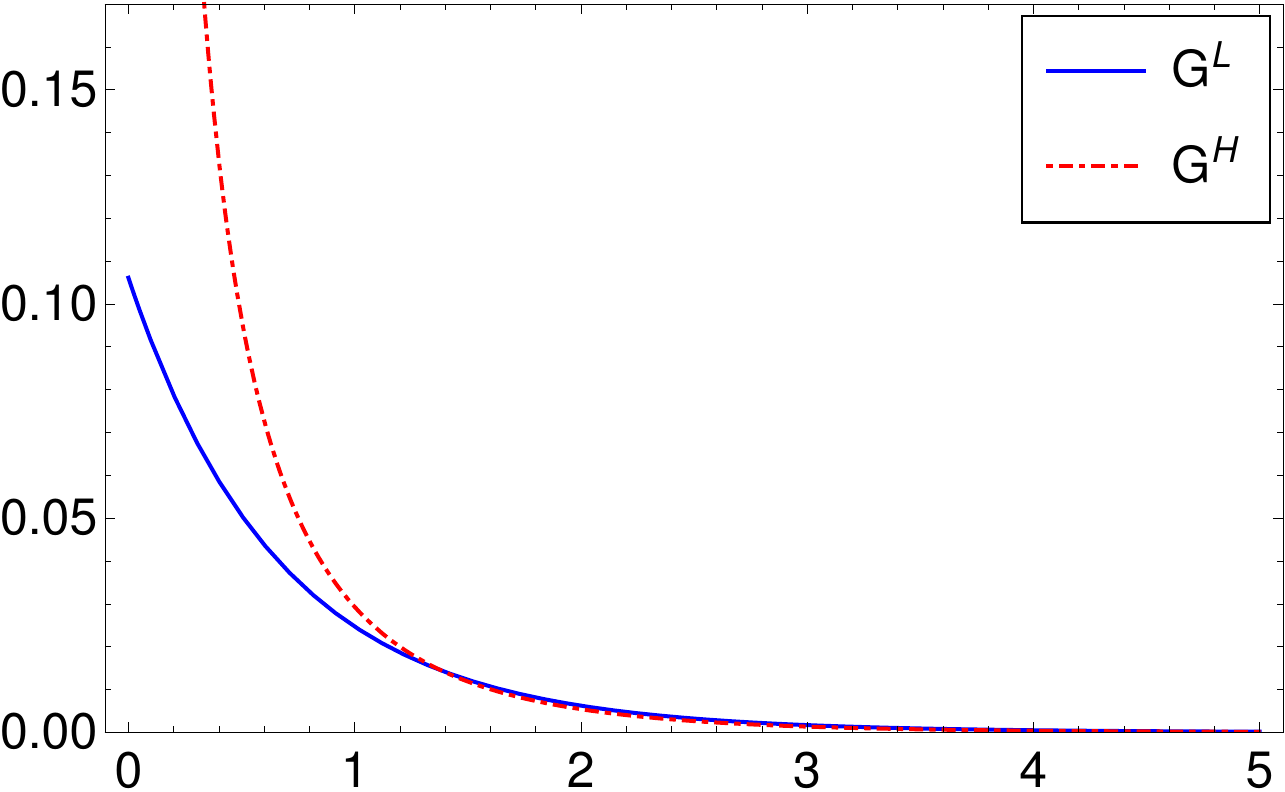,width=7.8cm}
\put(-4.0,-0.4){$R/a_1$}
}
\caption{Plot of the Green function $G^L$ for
 $a_1=2a_2$ in second gradient electromagnetostatics
and of the Green function $G^\text{H}$.}
\label{fig:GF}
\end{figure}

On the other hand, the Green function~\eqref{G-BKG-3d} is nonsingular and possesses a maximum 
value at $R=0$, namely (see  Fig.~\ref{fig:GF})
\begin{align}
\label{G-BH-0}
G^{L}(0)=\frac{1}{4\pi a_1 a_2 (a_1+a_2)}\,.
\end{align}
Moreover, the Green function~\eqref{G-BKG-3d} is a Dirac-delta sequence 
with parametric dependence $a_1$ and $a_2$
\begin{align}
\lim_{a_1\to 0}\lim_{a_2 \to 0} G^{L}(R)=
\lim_{a_1 \to 0} G^{\text{H}}(R)=\delta(\BR)
\end{align}
with 
\begin{align}
\lim_{a_2 \to 0} G^{L}(R)=G^{\text{H}}(R)\,,
\end{align}
where the limit is to be understood in the weak sense for distributions.
The Green functions \eqref{G-KG-3d} and \eqref{G-BKG-3d}
are plotted in Fig.~\ref{fig:GF}.

\subsection{Derivatives of the Green function $G^{L\Delta}$}
Now, we calculate the first, second and third  gradients 
of the Green function $G^{L\Delta}$.
The first, second and third gradients
of the Green function~\eqref{G-BP-3d} are obtained as\footnote{The expression $\text{sym}  (\bm 1 \bm R)$ means 
$\delta_{ij} R_k+ \delta_{jk} R_i+\delta_{ki} R_j$.}
\begin{align}
\label{G-BP-grad}
\nabla G^{L\Delta}(\bm R)
&=\frac{1}{4\pi}\, \frac{\bm R}{R^3}\,
f_1(R,a_1,a_2)\,,\\
\label{G-BP-grad2}
\nabla\nabla G^{L\Delta}(\bm R)
&=\frac{1}{4\pi}\, 
\Big[\frac{\bm 1}{R^3}\, f_1(R,a_1,a_2)-\frac{3\bm R\bm R}{R^5}\, f_2(R,a_1,a_2)
\Big]\,,\\
\label{G-BP-grad3}
\nabla\nabla\nabla G^{L\Delta}(\bm R)
&=-\frac{1}{4\pi}\, 
\bigg[\frac{3\, \text{sym}  (\bm 1 \bm R)}{R^5}\, f_2(R,a_1,a_2)-\frac{15\bm R\bm R\bm R}{R^7}\, f_3(R,a_1,a_2)
\bigg]
\end{align}
and (cf. Eq.~\eqref{BPE-3})
\begin{align}
\label{G-BP-Lapl}
\Delta G^{L\Delta}(\bm R)
&=\frac{3}{4\pi}\, 
\frac{1}{R^3}\,\big[ f_1(R,a_1,a_2)-f_2(R,a_1,a_2)\big]
\equiv
G^L(R)
\,,\\
\label{G-BP-Lapl-grad}
\nabla\Delta G^{L\Delta}(\bm R)
&=-\frac{15}{4\pi}\, 
\frac{\bm R}{R^5}\, \big[ f_2(R,a_1,a_2)- f_3(R,a_1,a_2)\big]
\equiv \nabla G^L(R)
\end{align}
with the auxiliary functions
\begin{align}
\label{f1}
f_1(R,a_1,a_2)=1
&-\frac{1}{a_1^2-a_2^2}\,
\Big[a_1^2 \e^{-R/a_1}-a_2^2 \e^{-R/a_2}\Big]
-\frac{R}{a_1^2-a_2^2}\,
\Big[a_1 \e^{-R/a_1}-a_2 \e^{-R/a_2}\Big]
\,,\\
\label{f2}
f_2(R,a_1,a_2)=1
&-\frac{1}{a_1^2-a_2^2}\,
\Big[a_1^2 \e^{-R/a_1}-a_2^2 \e^{-R/a_2}\Big]
-\frac{R}{a_1^2-a_2^2}\,
\Big[a_1 \e^{-R/a_1}-a_2 \e^{-R/a_2}\Big]
\nonumber\\
&-\frac{R^2}{3(a_1^2-a_2^2)}\,
\Big[\e^{-R/a_1}-\e^{-R/a_2}\Big]\,,\\
\label{f3}
f_3(R,a_1,a_2)=1
&-\frac{1}{a_1^2-a_2^2}\,\Big[a_1^2\,\e^{-R/a_1}-a_2^2\,\e^{-R/a_2}\Big]
-\frac{R}{a_1^2-a_2^2}\,\Big[a_1\,\e^{-R/a_1}-a_2\,\e^{-R/a_2}\Big]\nonumber\\
&-\frac{2R^2}{5(a_1^2-a_2^2)}\,\Big[\e^{-R/a_1}-\e^{-R/a_2}\Big]
-\frac{R^3}{15(a_1^2-a_2^2)}\,\Big[\frac{1}{a_1}\,
  \e^{-R/a_1}-\frac{1}{a_2}\,\e^{-R/a_2}\Big]\,.
\end{align}
Note that the auxiliary functions for the gradients of a Green function in the form \eqref{G-BP-2} obey
\begin{align}
\label{fi-relation}
 f_{i+1}(x, a_1, a_2) = f_i(x, a_1, a_2) - \frac{1}{2i + 1} \, x f_i'(x, a_1, a_2)
\end{align}
for $i = 0, 1, 2$.\footnote{This procedure might even carry on to higher orders of the derivatives in an approach to Eq.~\eqref{G-BP-2} similar to the results of 
\citep{Est-Kan1985}, but this level of generality is not needed here.}
This relation has direct consequences for the series expansions (the $f_i$ are analytic and can be differentiated term by term): $f_1$ has no linear term, $f_2$ neither linear nor cubic, $f_3$ no $R$, $R^3$ and $R^5$-term, and so forth.
Therefore, the first non-vanishing term of even order in $f_0$, which here is the fourth-order term, determines the strength of the regularization.

The relevant series expansions of the auxiliary functions~\eqref{f1}--\eqref{f3}
(near fields) read as
\begin{align}
\label{f1-ser}
f_1(R,a_1,a_2)&=\frac{1}{3 a_1a_2 (a_1+a_2)}\, R^3-\frac{1}{8 a^2_1a^2_2}\, R^4+\mathcal{O}(R^5)\,,\\
\label{f2-ser}
f_2(R,a_1,a_2)&=\frac{1}{24 a^2_1a^2_2}\, R^4 +\mathcal{O}(R^5)\,,\\
\label{f3-ser}
f_3(R,a_1,a_2)&=\frac{1}{120 a^2_1a^2_2}\, R^4 +\mathcal{O}(R^6)\,.
\end{align}
In Eqs.~\eqref{f1-ser}--\eqref{f3-ser},
it can be seen that the function $f_1(R,a_1,a_2)$ regularizes up to a
$1/R^3$-singularity and the functions $f_2(R,a_1,a_2)$ and  $f_3(R,a_1,a_2)$
regularize up to a $1/R^4$-singularity towards nonsingular field expressions.

\begin{figure}[t]\unitlength1cm
\vspace*{0.1cm}
\centerline{
\epsfig{figure=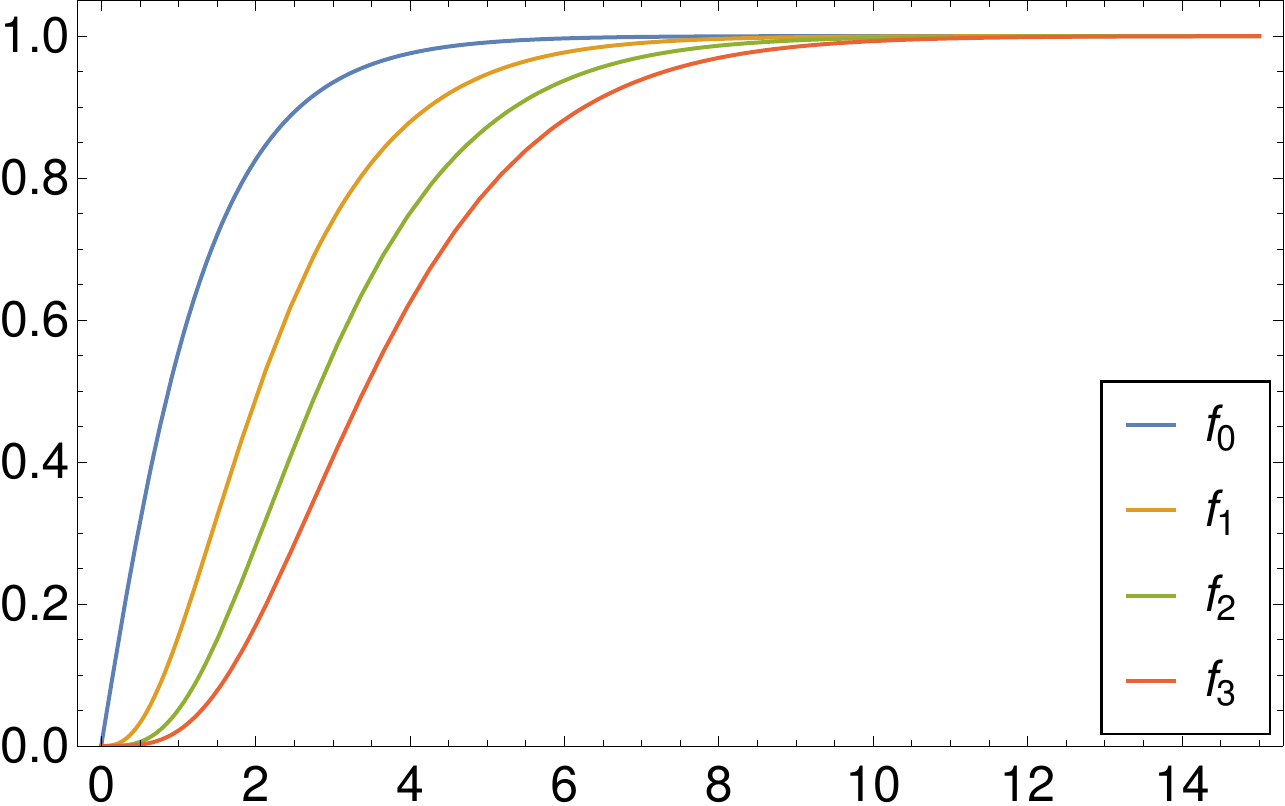,width=7.8cm}
\put(-4.0,-0.4){$R/a_1$}
}
\caption{Plot of the auxiliary functions $f_0$, $f_1$, $f_2$ and $f_3$ for
  $a_1=2a_2$ in second gradient electromagnetostatics.}
\label{fig:f0-3}
\end{figure}

The auxiliary functions~\eqref{f0}, \eqref{f1}--\eqref{f3}
are plotted in Fig.~\ref{fig:f0-3}. 
In the far field, the auxiliary functions~\eqref{f0}, \eqref{f1}--\eqref{f3}
approach the value of 1 and 
in the near field, they are modified due to gradient parts 
and approach the value of 0 at the position $R=0$ (see Fig.~\ref{fig:f0-3}).

\section{Electromagnetic fields
in second gradient electromagnetostatics}
\label{sec4}

The electromagnetic potentials are the solutions of the inhomogeneous partial differential equations of sixth order~\eqref{phi-w} and \eqref{A-w} 
for given charge and current densities ($\rho$, $\bm J$) 
\begin{align}
\label{phi-rp}
\phi&=-\frac{1}{\varepsilon_0}\, G^{L\Delta}*\rho\,,\\
\label{A-rp}
\bm A&=-\mu_0\, G^{L\Delta}*\bm J\,.
\end{align}

\subsection{Electric point charge}

The charge density of an electric point charge located at the position $\bm r'$ is given by
\begin{align}
\label{epc}
\rho=q\, \delta(\bm r-\bm r')\,,
\end{align}
where $q$ denotes the electric charge. This means that $\bm r'$ is the position vector of the point charge and $\bm r$ is the field vector.

\begin{figure}[t!!!]\unitlength1cm
\vspace*{0.1cm}
\centerline{
\epsfig{figure=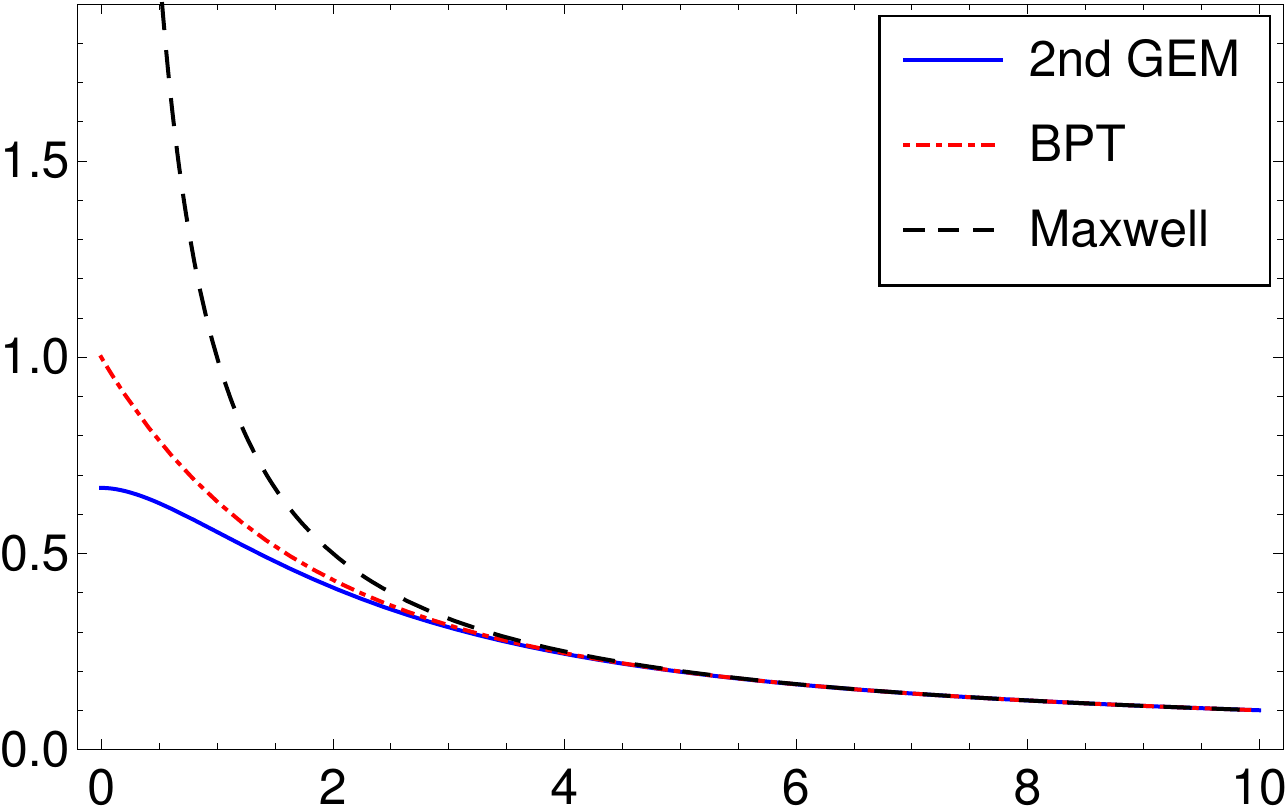,width=7.5cm}
\put(-3.9,-0.4){$R/a_1$}
\put(-7.6,-0.3){$\text{(a)}$}
\hspace*{0.2cm}
\put(-0.1,-0.3){$\text{(b)}$}
\put(3.6,-0.4){$R/a_1$}
\epsfig{figure=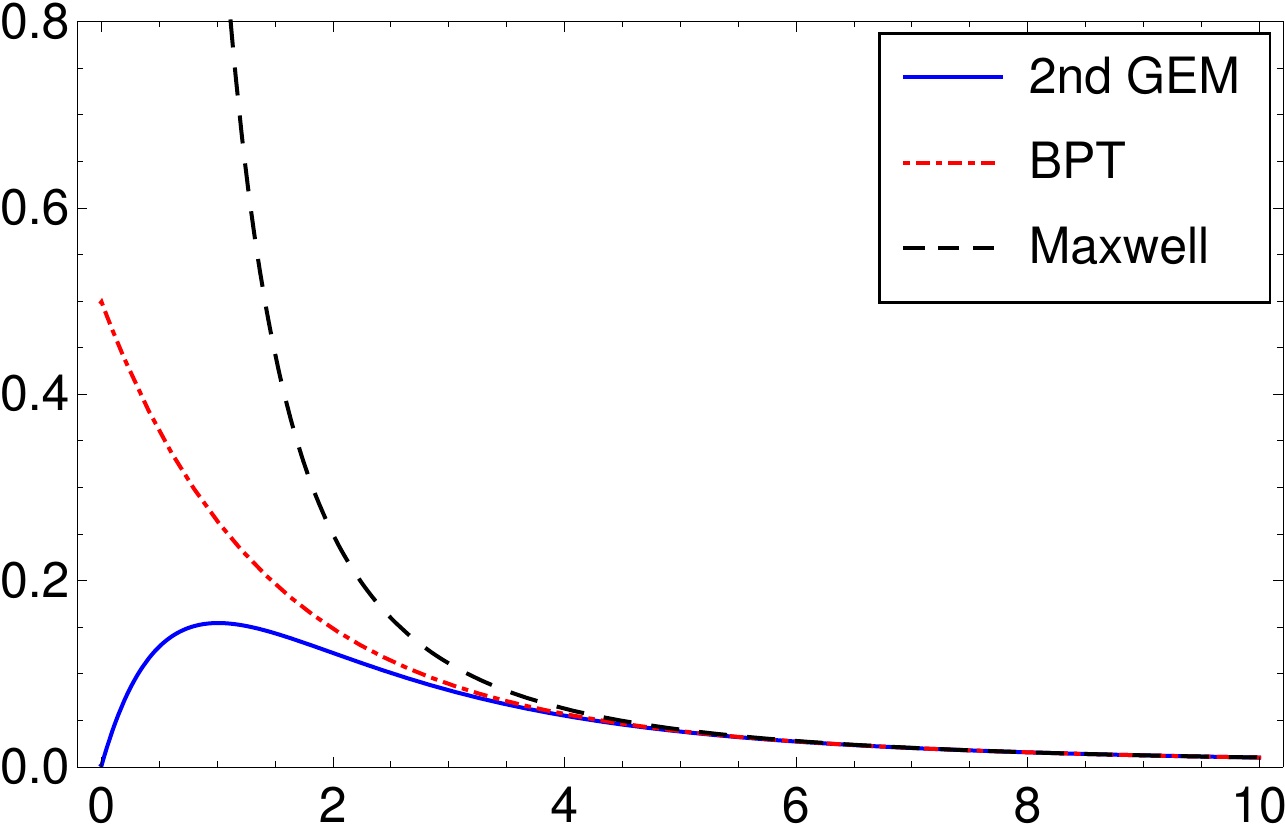,width=7.5cm}
}
\centerline{
\epsfig{figure=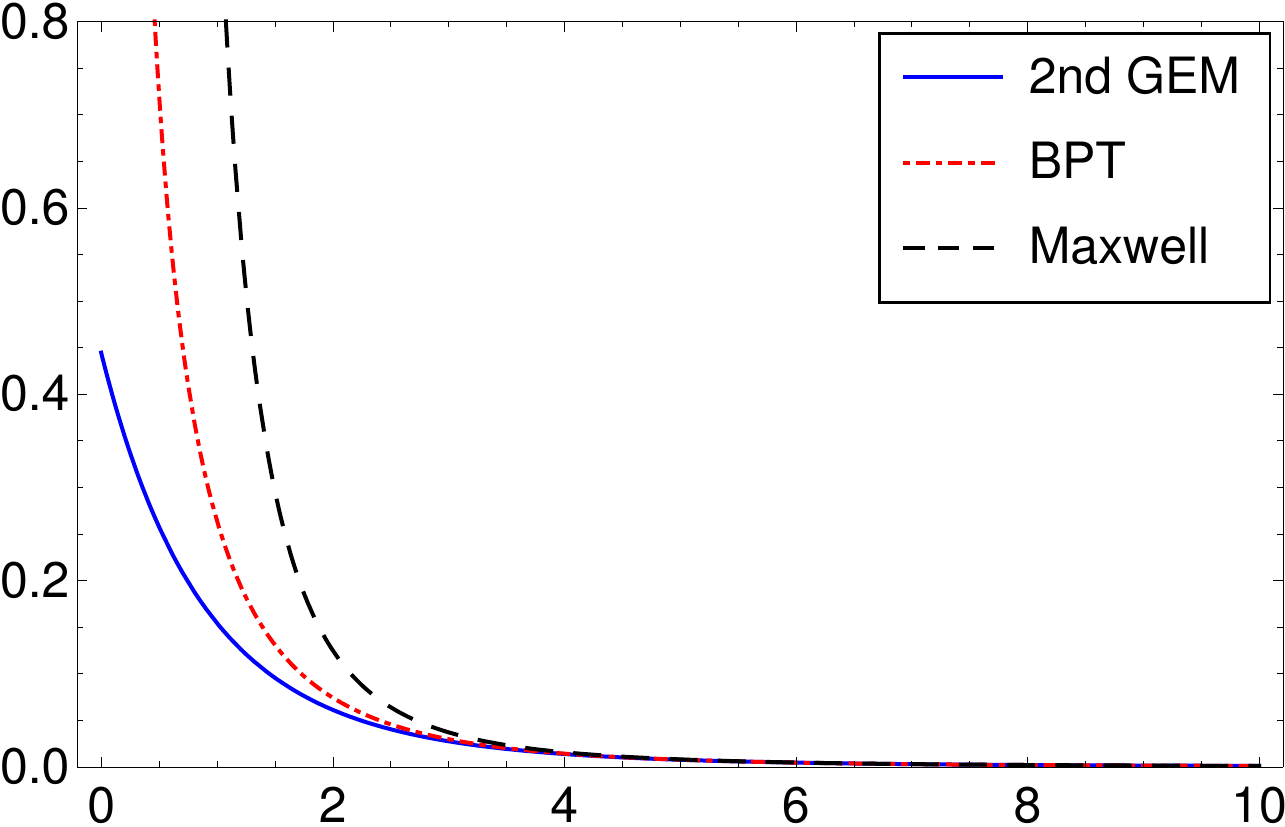,width=7.5cm}
\put(-3.9,-0.4){$R/a_1$}
\put(-7.6,-0.3){$\text{(c)}$}
\hspace*{0.2cm}
\put(3.6,-0.4){$R/a_1$}
\put(-0.1,-0.3){$\text{(d)}$}
\epsfig{figure=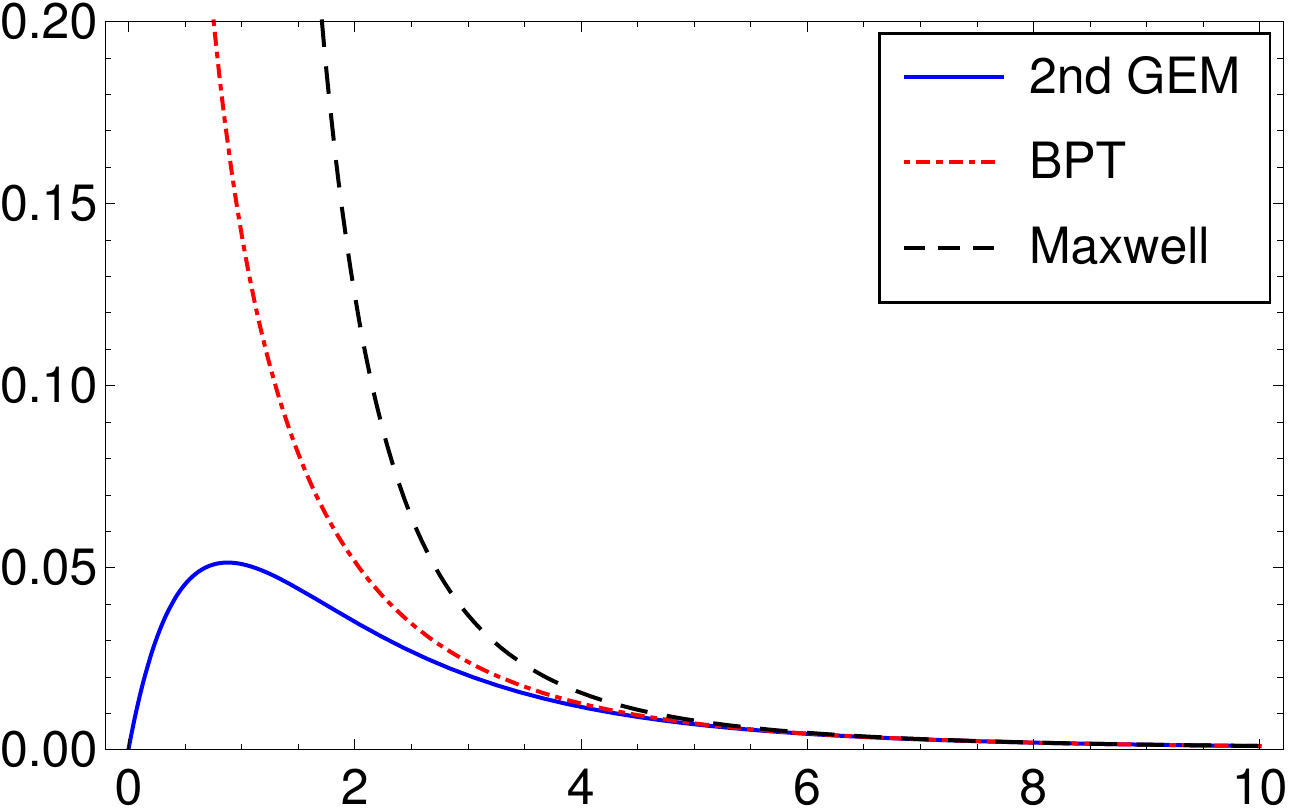,width=7.5cm}
}
\centerline{
\epsfig{figure=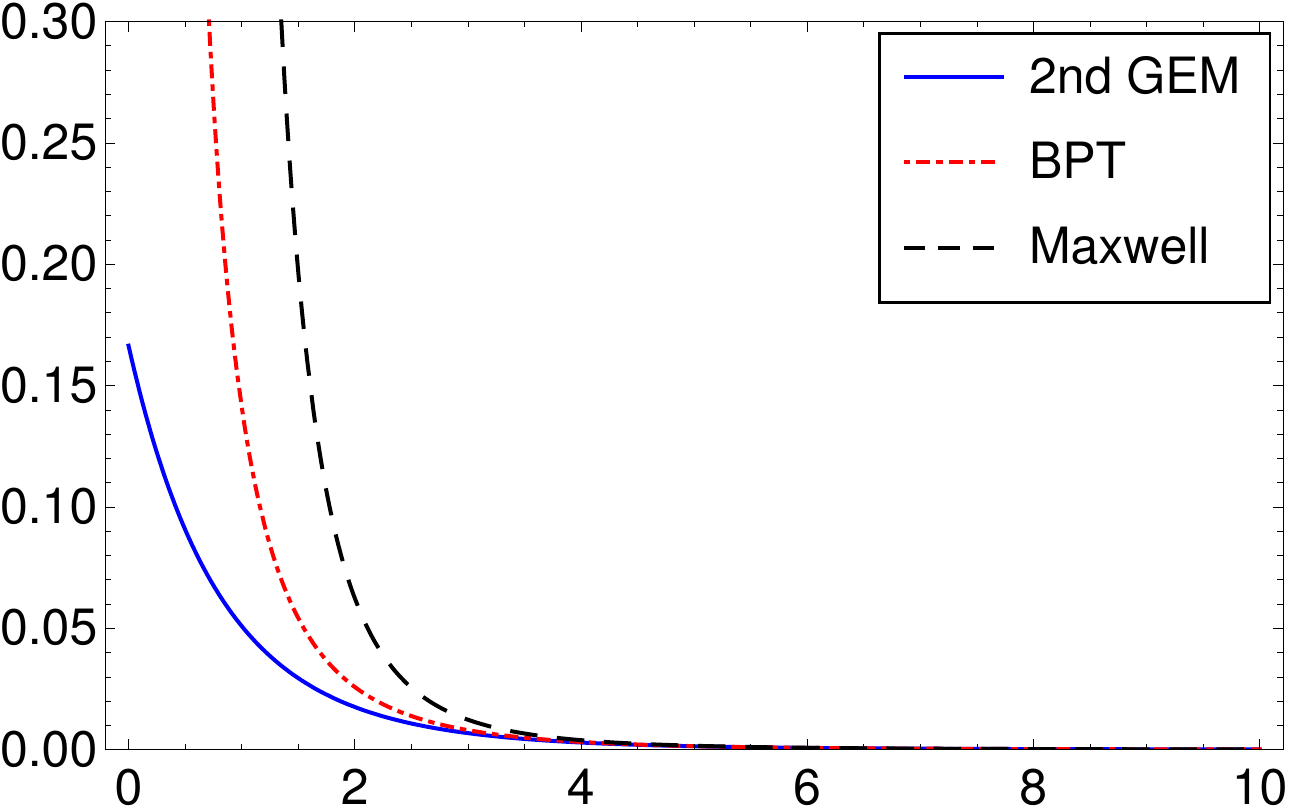,width=7.5cm}
\put(-3.9,-0.4){$R/a_1$}
\put(-7.6,-0.3){$\text{(e)}$}
\hspace*{0.2cm}
\put(3.6,-0.4){$R/a_1$}
\put(-0.1,-0.3){$\text{(f)}$}
\epsfig{figure=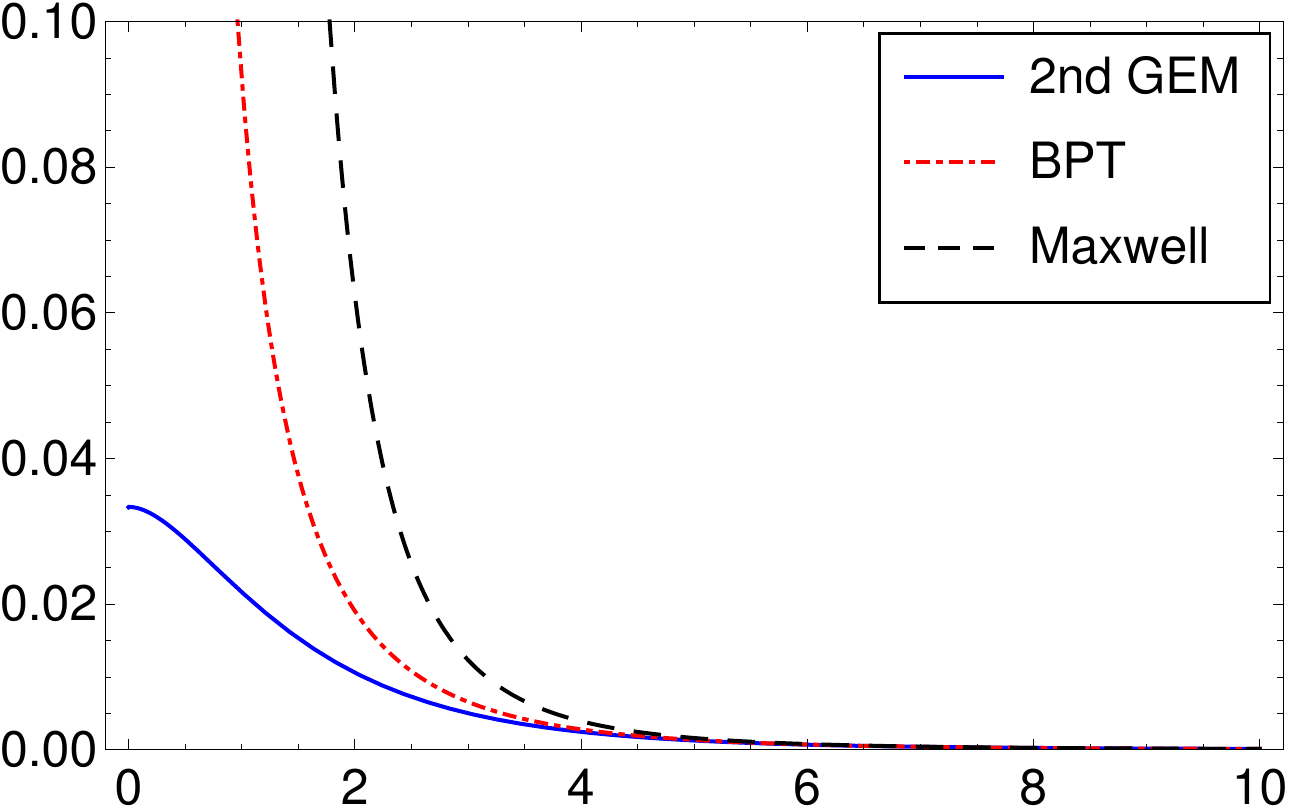,width=7.5cm}
}
\caption{Plots of the radial functions
in second gradient electromagnetostatics (2nd GEM) for $a_1=2a_2$,
Bopp-Podolsky electromagnetostatics (BPT) 
and classical Maxwell electromagnetostatics (Maxwell):
(a) $f_0/R$, (b) $f_1/R^2$, (c) $f_1/R^3$, (d) $f_2/R^3$,
(e) $f_2/R^4$ and (f) $f_3/R^4$.}
\label{fig:f}
\end{figure}

Substituting Eq.~\eqref{epc} into Eq.~\eqref{phi-rp} and performing the convolution, the electrostatic potential of a point charge reads as
\begin{align}
\label{phi-epc1}
\phi=-\frac{q}{\varepsilon_0}\, G^{L\Delta}\,.
\end{align}
If we insert the Green function~\eqref{G-BP-2} into Eq.~\eqref{phi-epc1}, the explicit expression of  the electrostatic potential of a point charge 
reads in terms of the auxiliary function~\eqref{f0}
\begin{align}
\label{phi-epc2}
\phi=\frac{q}{4\pi\varepsilon_0}\, \frac{1}{R}\, f_0(R,a_1,a_2)\,. 
\end{align}
Using the near field of $f_0$, Eq.~\eqref{f0-ser},  it can be seen that
the electrostatic potential~\eqref{phi-epc2} is finite at $R=0$, namely  (see Fig.~\ref{fig:f}a)
\begin{align} 
\label{phi-epc2-0}
\phi(0)=\frac{q}{4\pi\varepsilon_0(a_1+a_2)}\,. 
\end{align}

\begin{figure}[t!!!]\unitlength1cm
\vspace*{0.1cm}
\centerline{
\epsfig{figure=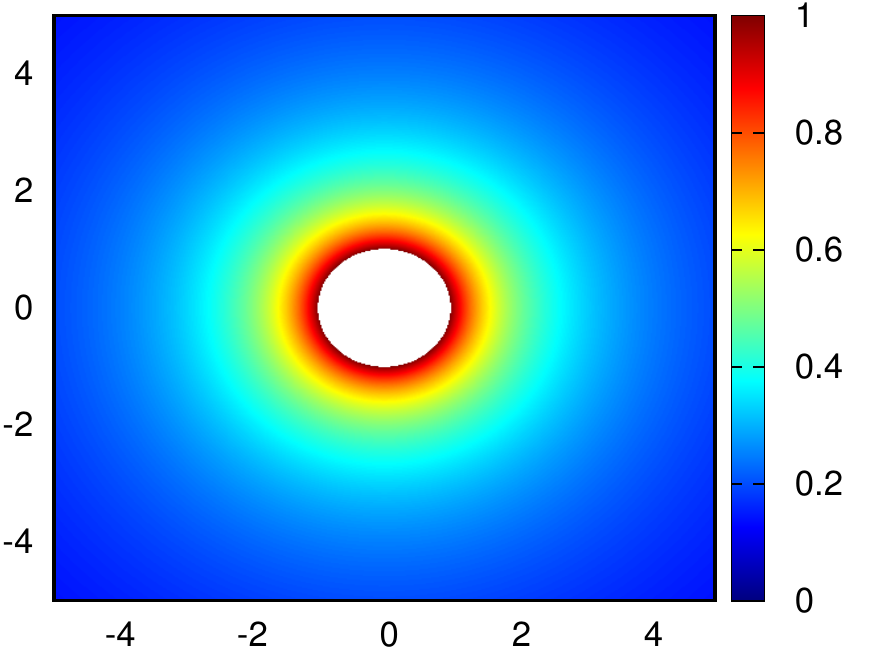,width=7.5cm}
\put(-4.35,-0.4){${\frac{x_1}{\ell}}$}
\put(-7.9,2.85){$\frac{x_3}{\ell}$}
\put(-7.6,-0.3){$\text{(a)}$}
\hspace*{0.2cm}
\put(-0.1,-0.3){$\text{(b)}$}
\put(3.15,-0.4){$\frac{x_1}{\ell}$}
\put(-0.4,2.85){$\frac{x_3}{\ell}$}
\epsfig{figure=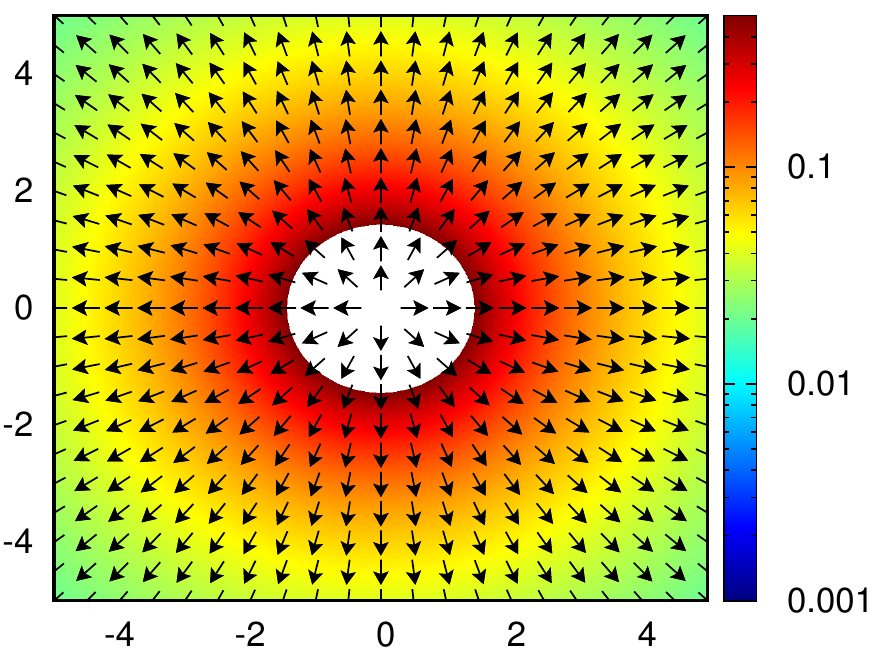,width=7.5cm}
}
\centerline{
\epsfig{figure=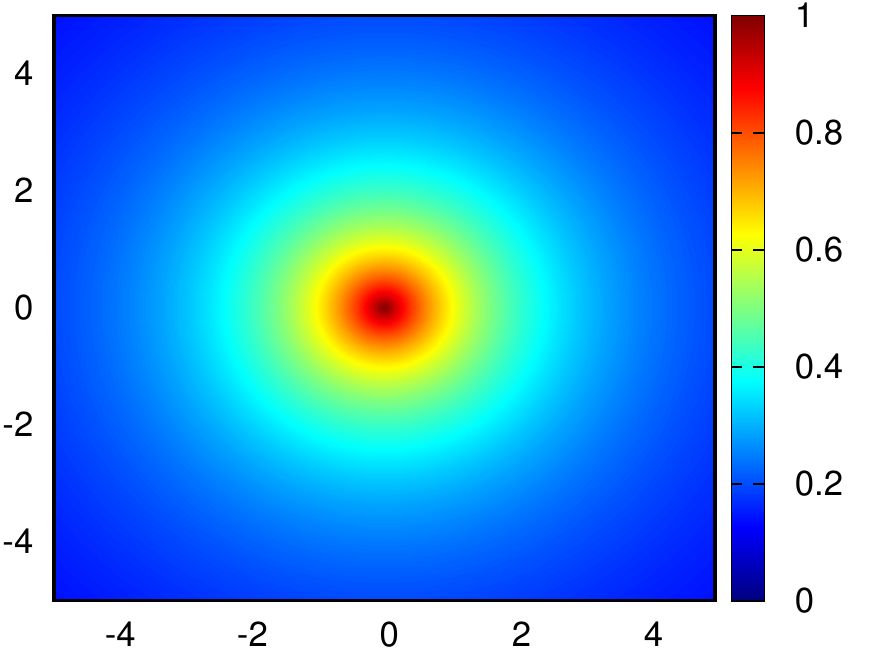,width=7.5cm}
\put(-4.35,-0.4){$\frac{x_1}{\ell}$}
\put(-7.9,2.85){$\frac{x_3}{\ell}$}
\put(-7.6,-0.3){$\text{(c)}$}
\hspace*{0.2cm}
\put(3.15,-0.4){$\frac{x_1}{\ell}$}
\put(-0.4,2.85){$\frac{x_3}{\ell}$}
\put(-0.1,-0.3){$\text{(d)}$}
\epsfig{figure=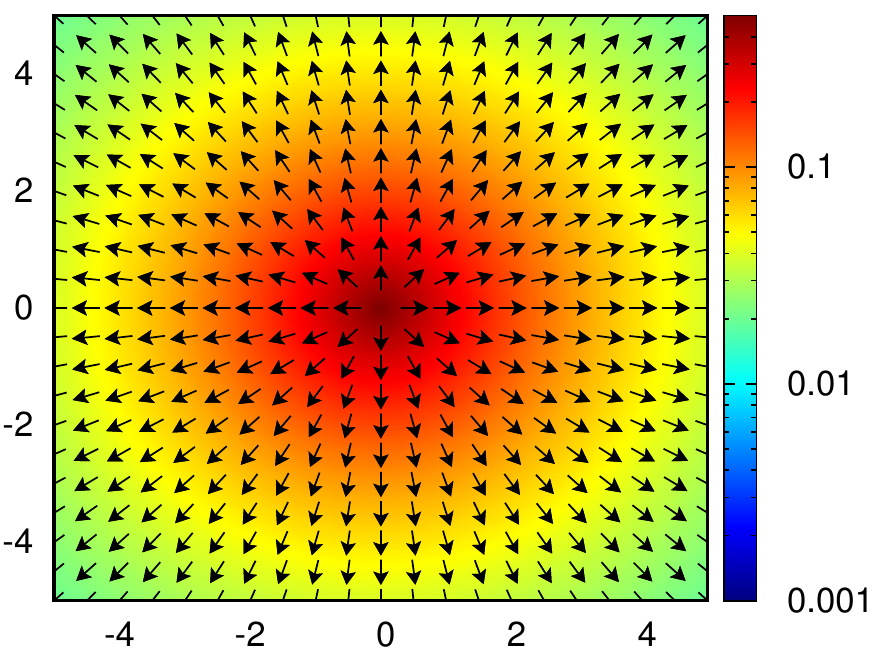,width=7.5cm}
}
\centerline{
\epsfig{figure=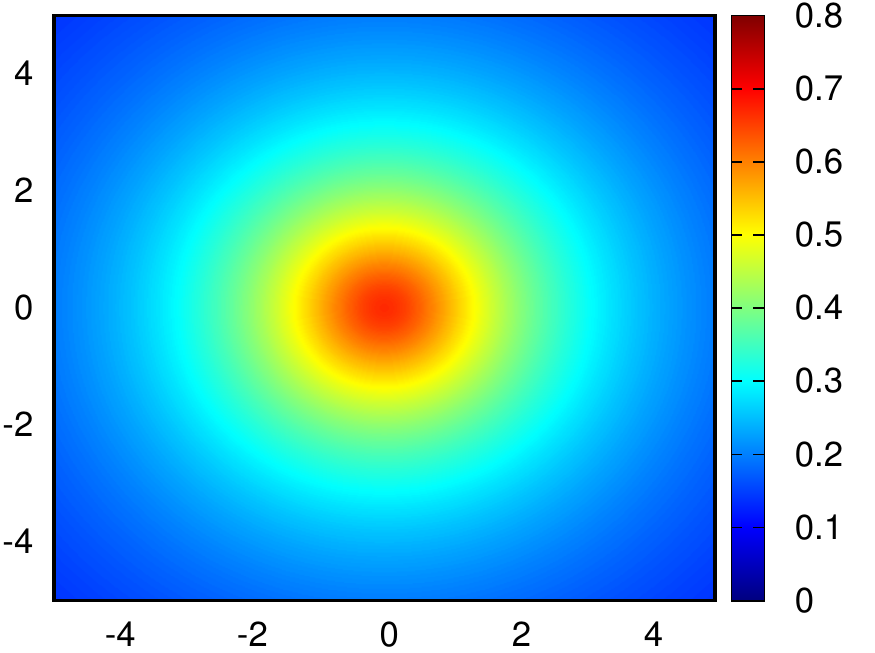,width=7.5cm}
\put(-4.35,-0.4){$\frac{x_1}{a_1}$}
\put(-7.9,2.85){$\frac{x_3}{a_1}$}
\put(-7.6,-0.3){$\text{(e)}$}
\hspace*{0.2cm}
\put(3.15,-0.4){$\frac{x_1}{a_1}$}
\put(-0.4,2.85){$\frac{x_3}{a_1}$}
\put(-0.1,-0.3){$\text{(f)}$}
\epsfig{figure=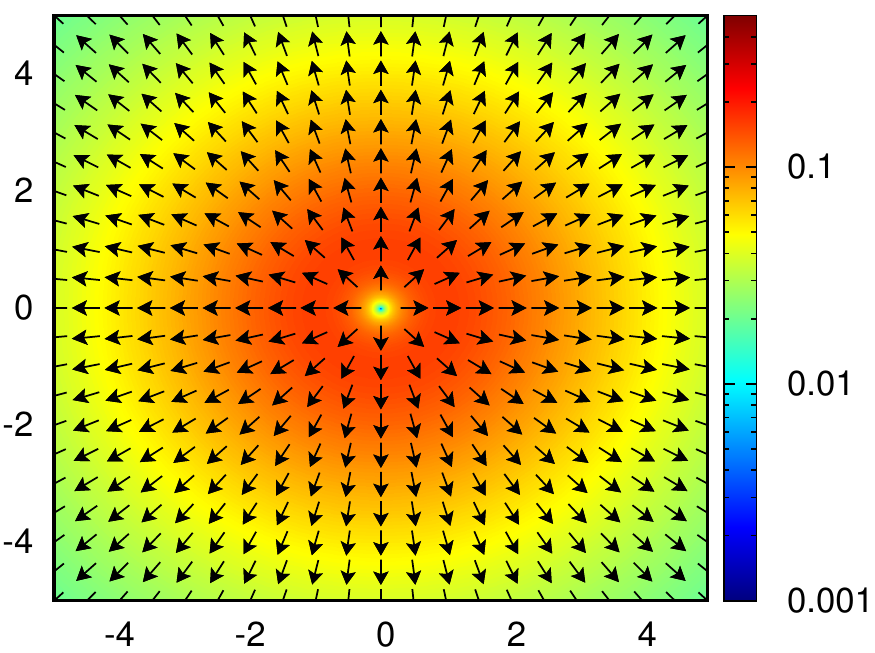,width=7.5cm}
}
\caption{Electric potential $\phi\,\frac{4 \pi \varepsilon_0 \ell}{q}$, (a) and (c), or $\phi\,\frac{4 \pi \varepsilon_0 a_1}{q}$, (e),  and electric field strength $\bm E\,\frac{4 \pi \varepsilon_0 \ell^2}{q}$, (b) and (d), or $\bm E\,\frac{4 \pi \varepsilon_0 a_1^2}{q}$, (f), in classical, (a) and (b), Bopp-Podolsky, (c) and (d), and second gradient electromagnetostatics, (e) and (f), for an electric point charge, for $x_2=0$.
 Arrows indicate field direction, the color its absolute value. Note that in (a) and (b) the color scale fails to display the singularities.}
\label{fig:e_pl}
\end{figure}
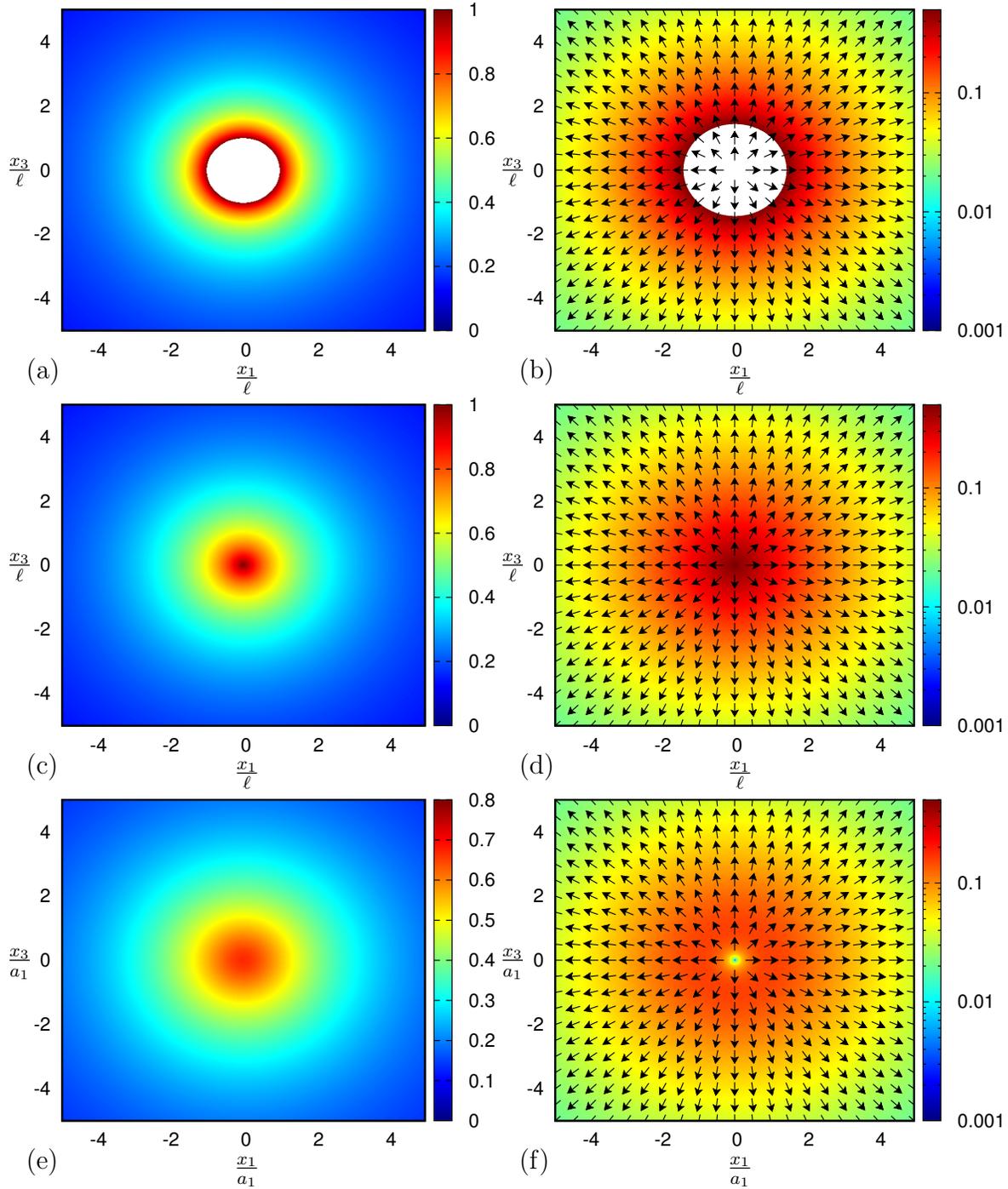

The electric field strength~\eqref{E} of a point charge is given by the negative gradient of the electrostatic potential~\eqref{phi-epc1}  and reads as
\begin{align}
\label{E-epc1}
\bm E=\frac{q}{\varepsilon_0}\, \nabla G^{L\Delta}\,.
\end{align}
Using Eq.~\eqref{G-BP-grad}, Eq.~\eqref{E-epc1} reduces to 
\begin{align}
\label{E-epc2}
\bm E=\frac{q}{4\pi\varepsilon_0}\, \frac{\bm R}{R^3}\, f_1(R,a_1,a_2)\,.
\end{align}
Using the near field of $f_1$, Eq.~\eqref{f1-ser},  it can be seen that
the electric field~\eqref{E-epc2} is zero at $R=0$.
In general, it is  nonsingular and possesses an extremum value  near the origin (see Fig.~\ref{fig:f}b). 
It does not have a directional discontinuity at the origin  unlike the electric field strength in the Bopp-Podolsky theory
(see Eq.~\eqref{E-epc2-BP} and Fig.~\ref{fig:f}b.)
For an overview and a comparison 
of point charge potentials and fields in second gradient electromagnetostatics and the limits to Bopp-Podolsky theory and classical theory see 
Fig.~\ref{fig:e_pl}. Unlike dipole fields, the electric field of the point charge in the gradient theory shows no changes in direction as compared to the classical result, it is, of course, spherically symmetric.

The electric field energy can be written as
\begin{align}
\label{Ue}
U_{\text{e}}&=\int_{\Bbb R^3}  \varepsilon_0
\Big(\bm E \cdot \bm E 
+\ell_1^2 \nabla \bm E :\nabla \bm E 
+\ell_2^4 \nabla\nabla \bm E \mathbin{\vdots} \nabla\nabla \bm E 
\Big)\text{d} V\nonumber\\
&=\int_{\Bbb R^3}  \varepsilon_0 \, L(\Delta) \bm E \cdot  \bm E\,  \text{d} V\nonumber\\
&=\int_{\Bbb R^3}   \bm D \cdot \bm E\,  \text{d} V\nonumber\\
&=\int_{\Bbb R^3}   \rho \,\phi\,  \text{d} V\,,
\end{align}
where we have used integration by parts and the surface terms vanish at infinity.
Substituting Eq.~\eqref{epc} into Eq.~\eqref{Ue},
the interaction energy  of two point charges reads 
\begin{align}
\label{Uqq}
U_{qq'}&=q \phi_{q'}\nonumber\\
&=\frac{q q'}{4\pi\varepsilon_0}\, \frac{1}{R}\, f_0(R,a_1,a_2)\,,
\end{align}
which is finite in the whole space. 
The electrostatic self-energy of a point charge is obtained as 
\begin{align}
\label{Uq}
U_{\text{self}}&=\frac{1}{2}\,U_{qq}(0)\nonumber\\
&=\frac{q^2}{8\pi\varepsilon_0\, (a_1+a_2)}\,,
\end{align}
which is finite.

The electrostatic part of the Lorentz force reads as~\citep{Lazar20}
\begin{align}
\label{F-L-e}
\bm{{F}}=\int_{\Bbb R^3} \rho \bm E \, \d V\,.
\end{align}
Substituting Eqs.~\eqref{epc} and \eqref{E-epc2} into Eq.~\eqref{F-L-e},
the electrostatic interaction force between two point charges $q'$ at $\bm r'$ and $q$ at $\bm r$ is obtained as
\begin{align}
\label{Fqq}
{\bm{F}}_{qq'}&=q \bm E_{q'}\nonumber\\
&=\frac{qq'}{4\pi\varepsilon_0}\, \frac{\bm R}{R^3}\, f_1(R,a_1,a_2)\,,
\end{align}
which is zero at $R=0$ and nonsingular. 
Eq.~\eqref{Fqq} is the force exerted by one charge $q'$ at $\bm r'$ on the other charge $q$ at $\bm r$.
It holds $\bm{F}_{qq'}=-\bm{F}_{q'q}$.

\subsection{Electric dipole}

The charge density of an ideal electric dipole is given by  
\begin{align}
\label{ed}
\rho=-\bm p\cdot \nabla\, \delta(\bm r-\bm r')\,,
\end{align}
where $\bm p$ is the electric dipole moment.

Substituting Eq.~\eqref{ed} into Eq.~\eqref{phi-rp} and employing the convolution, we obtain for  
the electrostatic potential of an electric dipole, or the electric dipole potential, 
\begin{align}
\label{phi-ed1}
\phi=\frac{1}{\varepsilon_0}\, \bm p\cdot \nabla\ G^{L\Delta}\,.
\end{align}
Now, using Eq.~\eqref{G-BP-grad}, Eq.~\eqref{phi-ed1} becomes
\begin{align}
\label{phi-ed2}
\phi=\frac{1}{4\pi\varepsilon_0}\, \frac{\bm p\cdot \bm R}{R^3}\, f_1(R,a_1,a_2)\,.
\end{align}
It is zero at $\bm R=0$ and possesses an extremum value  near the origin (see Fig.~\ref{fig:f}b).

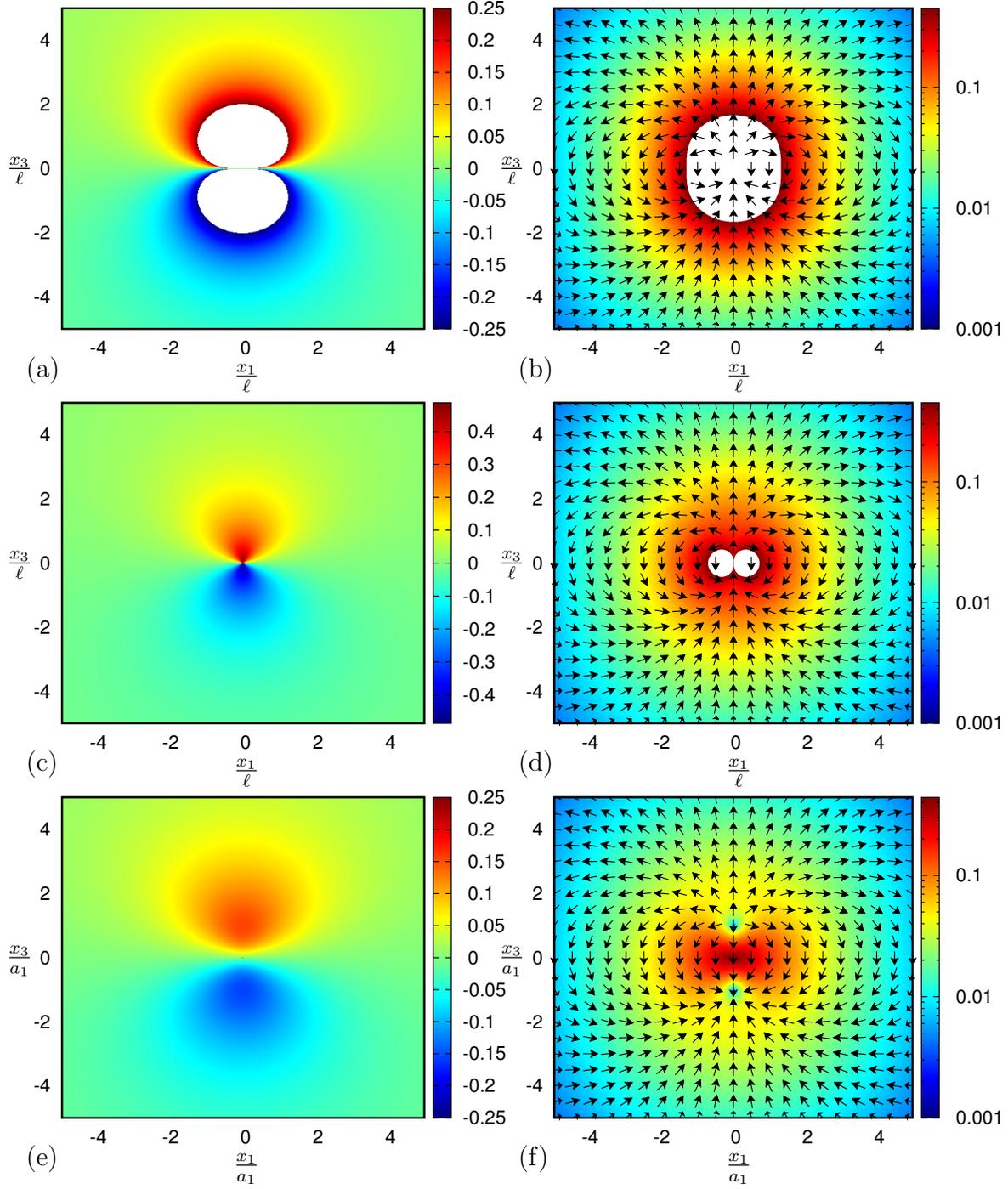
\begin{figure}[t!!!]\unitlength1cm
\centerline{
\epsfig{figure=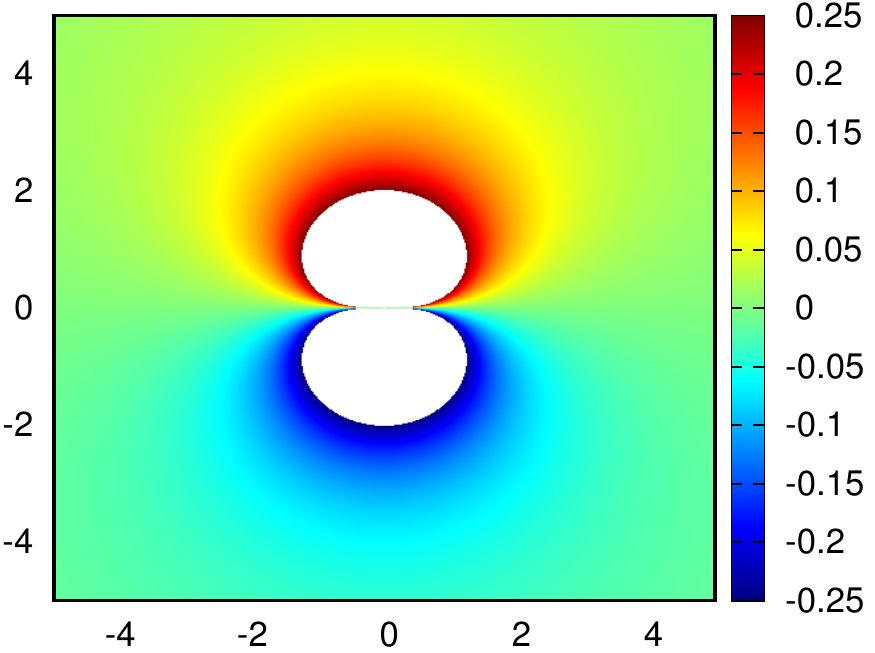,width=7.5cm}
\put(-4.35,-0.4){$\frac{x_1}{\ell}$}
\put(-7.9,2.85){$\frac{x_3}{\ell}$}
\put(-7.6,-0.3){$\text{(a)}$}
\hspace*{0.2cm}
\put(-0.1,-0.3){$\text{(b)}$}
\put(3.15,-0.4){$\frac{x_1}{\ell}$}
\put(-0.4,2.85){$\frac{x_3}{\ell}$}
\epsfig{figure=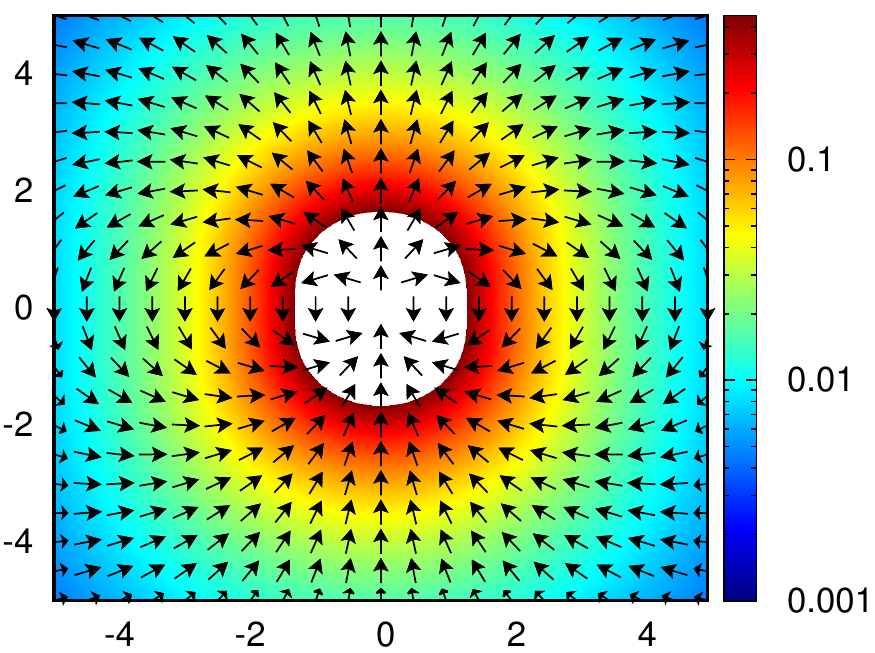,width=7.5cm}
}
\centerline{
\epsfig{figure=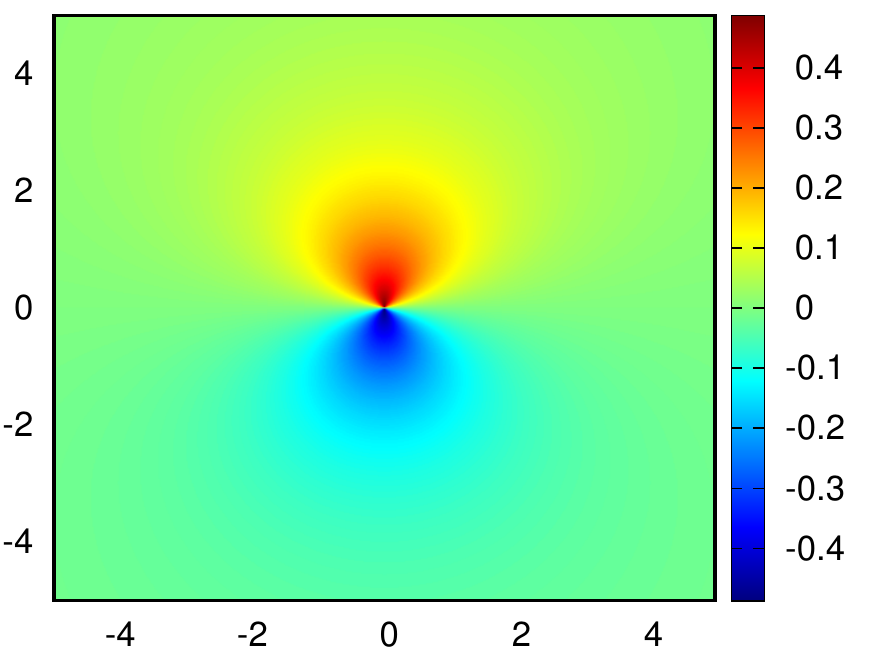,width=7.5cm}
\put(-4.35,-0.4){$\frac{x_1}{\ell}$}
\put(-7.9,2.85){$\frac{x_3}{\ell}$}
\put(-7.6,-0.3){$\text{(c)}$}
\hspace*{0.2cm}
\put(3.15,-0.4){$\frac{x_1}{\ell}$}
\put(-0.4,2.85){$\frac{x_3}{\ell}$}
\put(-0.1,-0.3){$\text{(d)}$}
\epsfig{figure=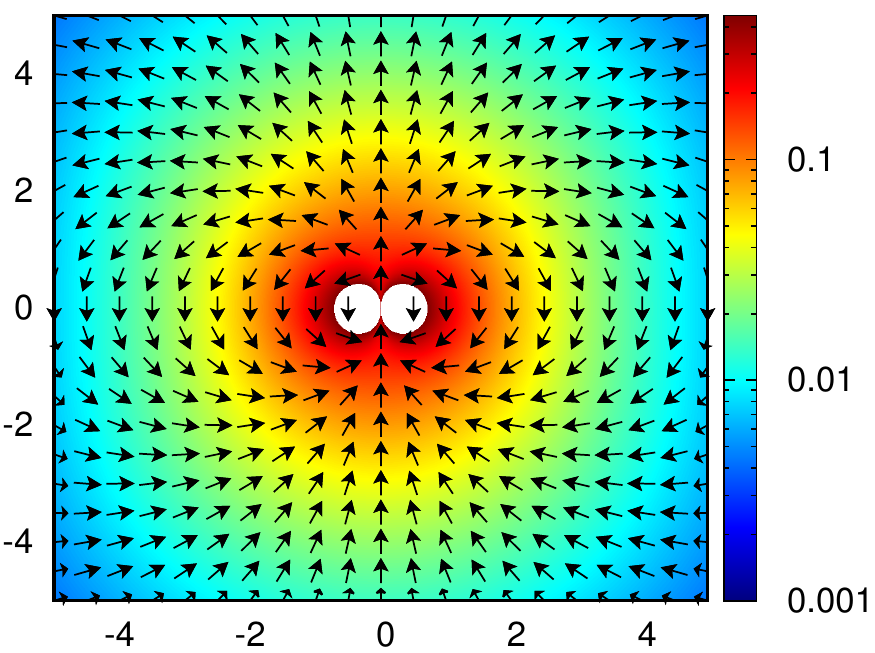,width=7.5cm}
}
\centerline{
\epsfig{figure=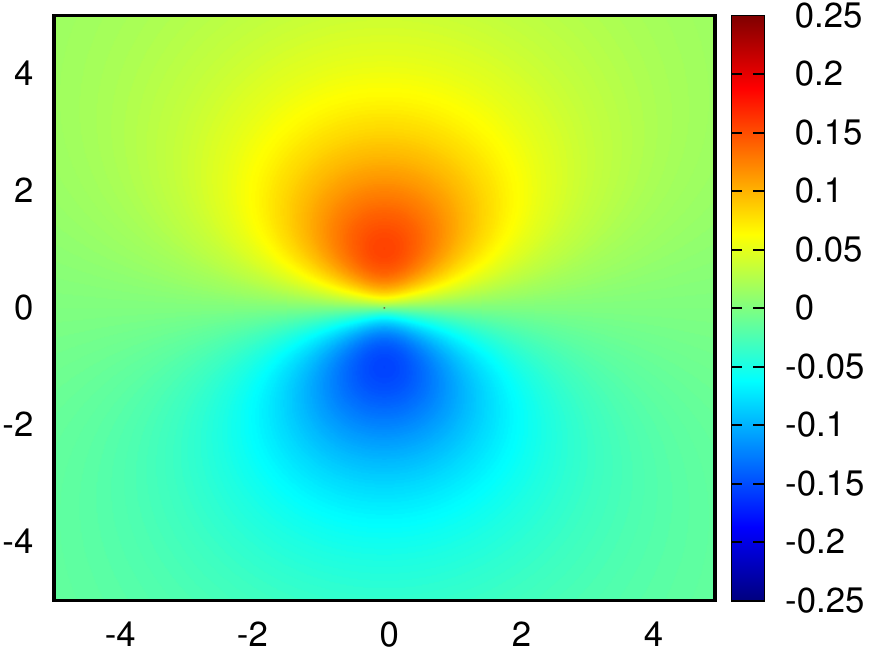,width=7.5cm}
\put(-4.35,-0.4){$\frac{x_1}{a_1}$}
\put(-7.9,2.85){$\frac{x_3}{a_1}$}
\put(-7.6,-0.3){$\text{(e)}$}
\hspace*{0.2cm}
\put(3.15,-0.4){$\frac{x_1}{a_1}$}
\put(-0.4,2.85){$\frac{x_3}{a_1}$}
\put(-0.1,-0.3){$\text{(f)}$}
\epsfig{figure=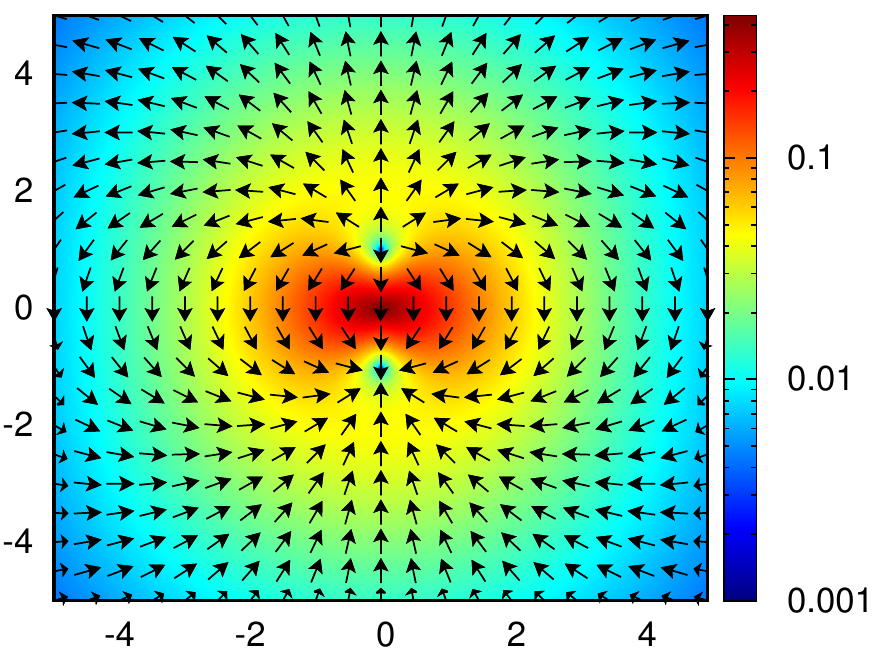,width=7.5cm}
}
\caption{Electric potential $\phi\,\frac{4 \pi \varepsilon_0 \ell^2}{p}$, (a) and (c), or $\phi\,\frac{4 \pi \varepsilon_0 a_1^2}{p}$, (e),  and electric field strength $\bm E\,\frac{4 \pi \varepsilon_0 \ell^3}{p}$, (b) and (d), or $\bm E\,\frac{4 \pi \varepsilon_0 a_1^3}{p}$, (f), in classical, (a) and (b), Bopp-Podolsky, (c) and (d), and second gradient electromagnetostatics, (e) and (f), for an electric dipole with $\bm p = p \bm e_3$, $\bm e_3$ being the unit vector along the third coordinate, for $x_2=0$. Arrows indicate field direction, the color its absolute value. Note that in (a), (b) and (d) the color scale fails to display the singularities. }
\label{fig:e_dip}
\end{figure}

The electric field strength~\eqref{E} of an electric dipole is given by the negative gradient of the electric dipole potential~\eqref{phi-ed1}
\begin{align}
\label{E-ed1}
\bm E=-\frac{1}{\varepsilon_0}\, (\bm p\cdot \nabla) \nabla G^{L\Delta}
\end{align}
and, using Eq~\eqref{G-BP-grad2}, it reads as
\begin{align}
\label{E-ed2}
\bm E=\frac{1}{4\pi\varepsilon_0}\, 
\bigg[
\frac{3(\bm p\cdot\bm R) \bm R}{R^5}\, f_2(R,a_1,a_2)
-\frac{\bm p}{R^3}\, f_1(R,a_1,a_2)
\bigg]\,.
\end{align}
Moreover, using the near fields of $f_1$ and $f_2$, Eqs.~\eqref{f1-ser} and \eqref{f2-ser},
it can be seen that
the electric dipole field~\eqref{E-ed2} is finite at $R=0$ due to the $f_1$-term, namely  (see Figs.~\ref{fig:f}c and \ref{fig:f}d)  
\begin{align}
\label{E-ed2-0}
\bm E(0)=-\frac{\bm p}{12\pi\varepsilon_0a_1a_2(a_1+a_2)}\,,
\end{align}
and it does not have a directional discontinuity. 
See Fig.~\ref{fig:e_dip} for a comparison of dipole potentials and fields in second gradient electromagnetostatics and the limits to Bopp-Podolsky theory and 
classical Maxwell theory. Along with the regularization directional changes in the electric field of the dipole in the gradient theory are introduced, due to the appearance of two different auxiliary functions in the two terms of the dipole field \eqref{E-ed2}. These modifications are stronger in second gradient electromagnetostatics than in the 
Bopp-Podolsky theory and even lead to qualitatively new behavior: There exist two zeros of the dipole field.

Substituting Eq.~\eqref{ed} into Eq.~\eqref{Ue},
the interaction energy of two electric dipoles $\bm p$ and $\bm p'$, a distance $\bm R=\bm r-\bm r'$ apart, 
becomes
\begin{align}
\label{Upp}
U_{\bm p \bm p '}&=-\bm p\cdot \bm E_{p'}\nonumber\\
&=
\frac{1}{4\pi\varepsilon_0}\, 
\bigg[\frac{\bm p\cdot\bm p'}{R^3}\, f_1(R,a_1,a_2)
-\frac{3(\bm p\cdot\bm R) (\bm p'\cdot\bm R)}{R^5}\, f_2(R,a_1,a_2)
\bigg]\,.
\end{align}
The electrostatic self-energy of an electric dipole reduces to 
\begin{align}
\label{Up}
U_{\text{self}}&=\frac{1}{2}\,U_{\bm p \bm p}(0)\nonumber\\
&=\frac{\bm p\cdot\bm p}{24\pi\varepsilon_0 a_1a_2(a_1+a_2)}\,,
\end{align}
which is finite. 

Substituting Eqs.~\eqref{ed} and \eqref{E-ed1} into Eq.~\eqref{F-L-e} and using Eq.~\eqref{G-BP-grad3},
the electrostatic interaction force between two electric dipoles reduces to 
\begin{align}
\label{Fpp}
{\bm {F}}_{\bm p \bm p '}&=(\bm p\cdot \nabla) \bm E_{p'}\nonumber\\
&=-\frac{1}{\varepsilon_0}\, (\bm p\cdot \nabla) (\bm p'\cdot \nabla) \, \nabla  G^{L\Delta}\nonumber\\
&=\frac{1}{4\pi\varepsilon_0}\,
\bigg[\frac{3(\bm p\cdot \bm R) \bm p'+3(\bm p'\cdot \bm R) \bm p + 3(\bm p\cdot \bm p') \bm R}{R^5}\, f_2(R,a_1,a_2)
\nonumber\\
&\qquad
-\frac{15(\bm p\cdot\bm R)(\bm p'\cdot\bm R) \bm R}{R^7}\, f_3(R,a_1,a_2)
\bigg]\,,
\end{align}
which is finite at $R=0$ and nonsingular (see Figs.~\ref{fig:f}e and \ref{fig:f}f), 
but it has a directional discontinuity at $R=0$. 
Eq.~\eqref{Fqq} is the force exerted by one electric dipole $p'$ at $\bm r'$ on the other electric dipole $p$ at $\bm r$.
It holds $\bm{F}_{pp'}=-\bm{F}_{p'p}$.

\subsection{Magnetic dipole}

The electric current density vector of a magnetic dipole is given by  
\begin{align}
\label{md}
\bm J=-\bm m\times \nabla\, \delta(\bm r-\bm r')\,,
\end{align}
where $\bm m$ is the magnetic dipole moment.

Substituting Eq.~\eqref{md} into Eq.~\eqref{A-rp} and 
performing the convolution, we obtain for  
the magnetic vector potential of a magnetic dipole, 
or the magnetic dipole potential, 
\begin{align}
\label{A-md1}
\bm A=\mu_0\, \bm m\times \nabla\, G^{L\Delta}\,.
\end{align}
Using Eq.~\eqref{G-BP-grad}, Eq.~\eqref{A-md1} becomes
\begin{align}
\label{A-md2}
\bm A=\frac{\mu_0}{4\pi}\, \frac{\bm m\times \bm R}{R^3}\, f_1(R,a_1,a_2)\,,
\end{align}
which is zero at $\bm R=0$ and possesses an extremum value  
(see Fig.~\ref{fig:f}b).

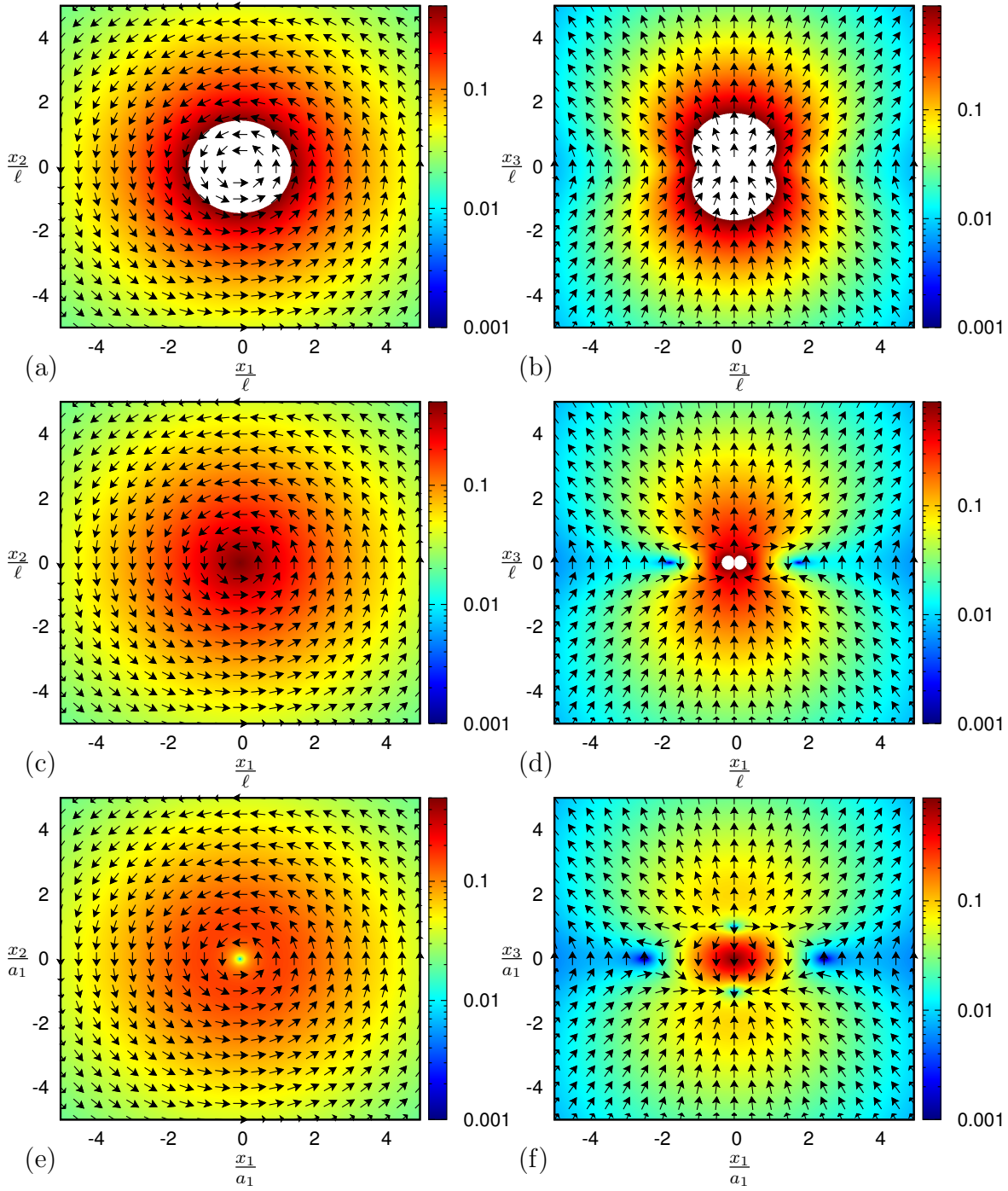
\begin{figure}[t!!!]\unitlength1cm
\centerline{
\epsfig{figure=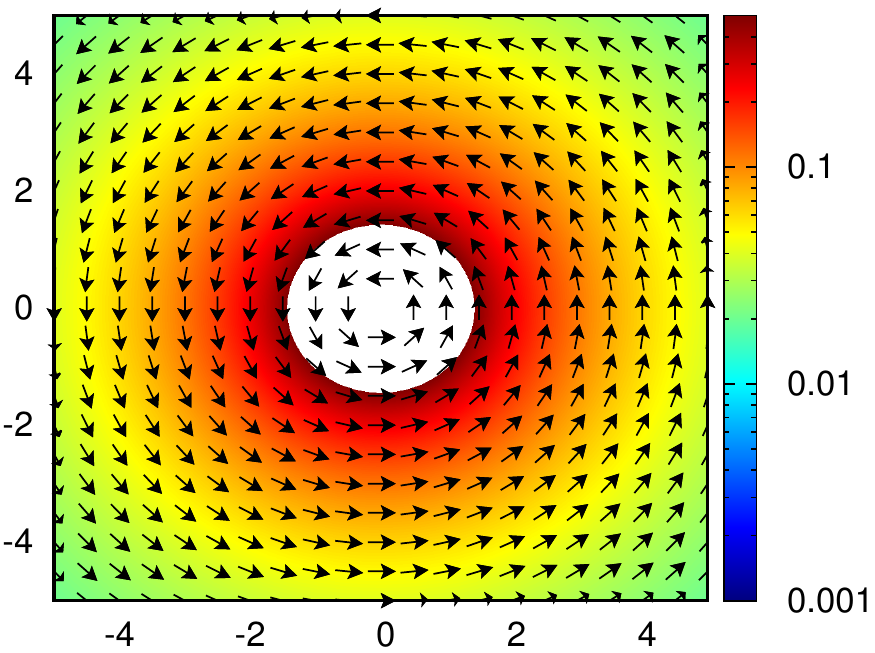,width=7.5cm}
\put(-4.35,-0.4){$\frac{x_1}{\ell}$}
\put(-7.9,2.85){$\frac{x_2}{\ell}$}
\put(-7.6,-0.3){$\text{(a)}$}
\hspace*{0.2cm}
\put(-0.1,-0.3){$\text{(b)}$}
\put(3.15,-0.4){$\frac{x_1}{\ell}$}
\put(-0.4,2.85){$\frac{x_3}{\ell}$}
\epsfig{figure=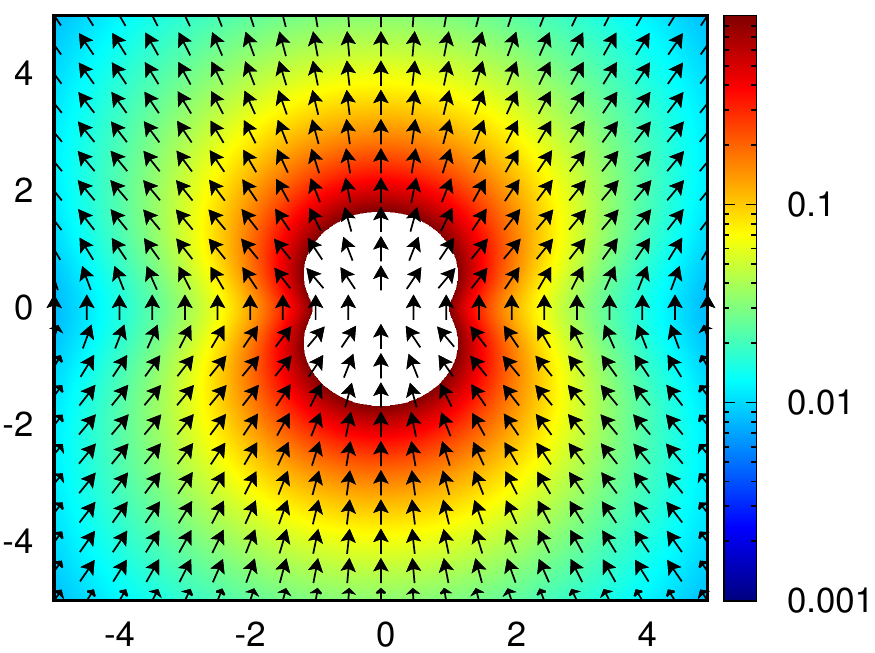,width=7.5cm}
}
\centerline{
\epsfig{figure=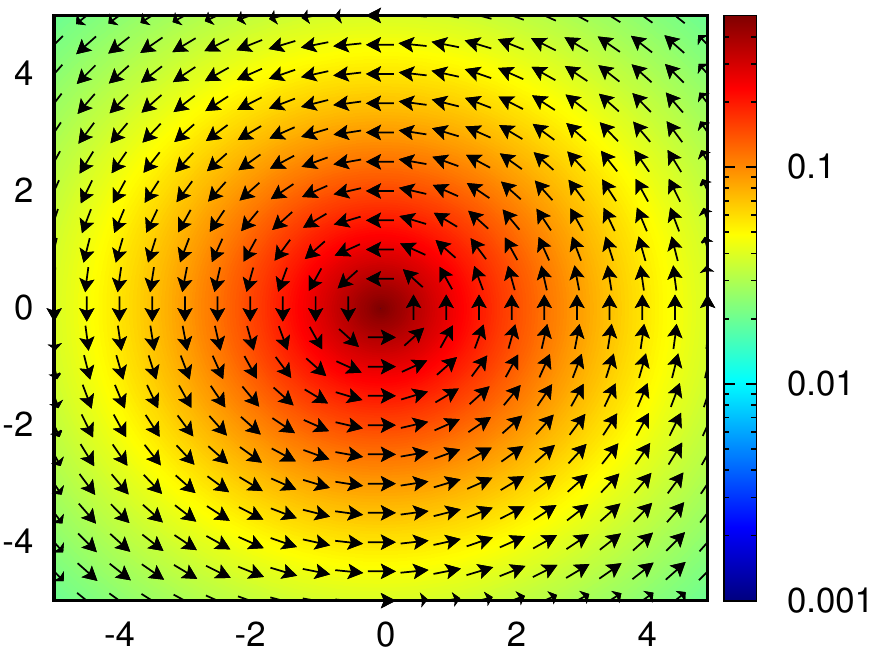,width=7.5cm}
\put(-4.35,-0.4){$\frac{x_1}{\ell}$}
\put(-7.9,2.85){$\frac{x_2}{\ell}$}
\put(-7.6,-0.3){$\text{(c)}$}
\hspace*{0.2cm}
\put(3.15,-0.4){$\frac{x_1}{\ell}$}
\put(-0.4,2.85){$\frac{x_3}{\ell}$}
\put(-0.1,-0.3){$\text{(d)}$}
\epsfig{figure=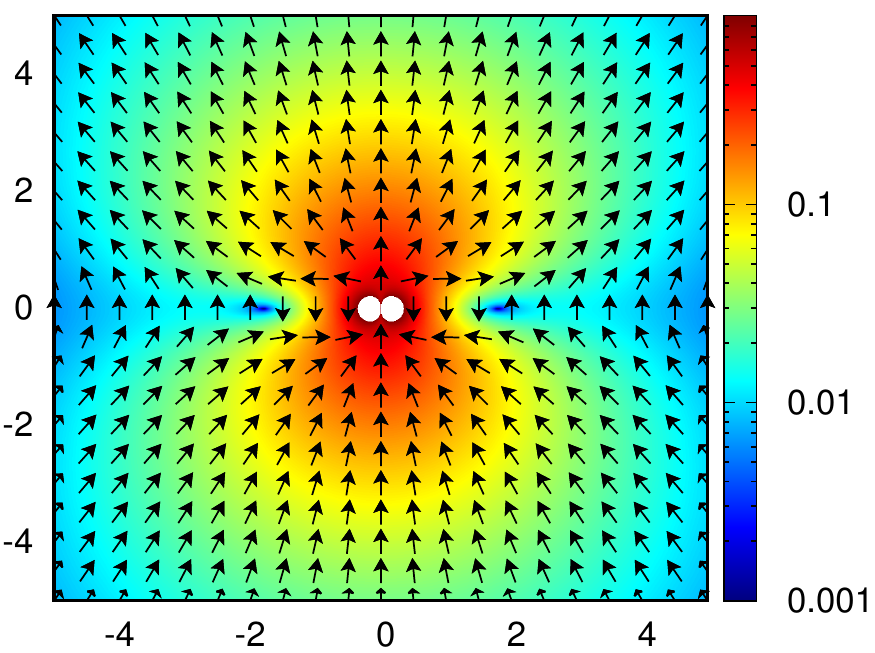,width=7.5cm}
}
\centerline{
\epsfig{figure=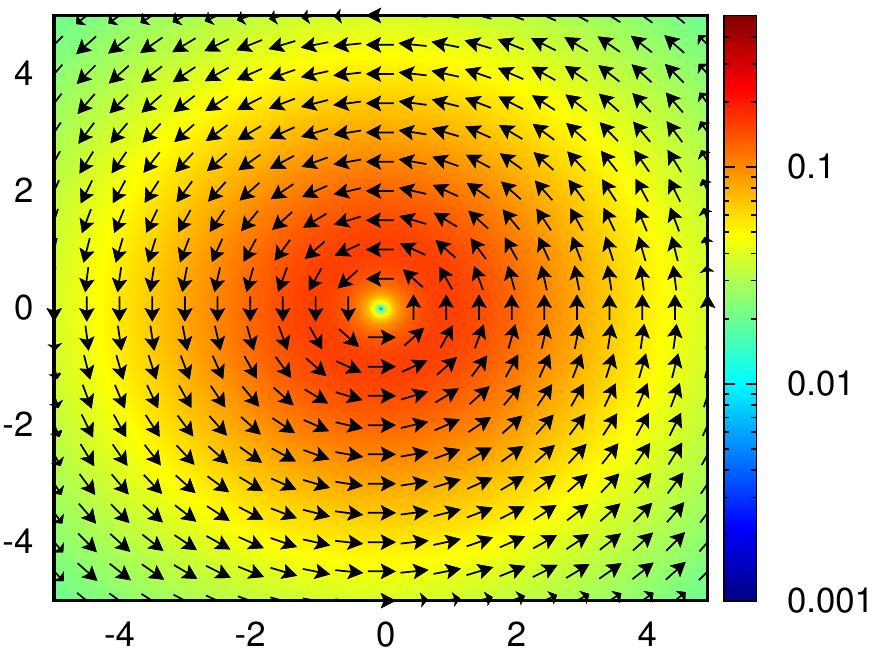,width=7.5cm}
\put(-4.35,-0.4){$\frac{x_1}{a_1}$}
\put(-7.9,2.85){$\frac{x_2}{a_1}$}
\put(-7.6,-0.3){$\text{(e)}$}
\hspace*{0.2cm}
\put(3.15,-0.4){$\frac{x_1}{a_1}$}
\put(-0.4,2.85){$\frac{x_3}{a_1}$}
\put(-0.1,-0.3){$\text{(f)}$}
\epsfig{figure=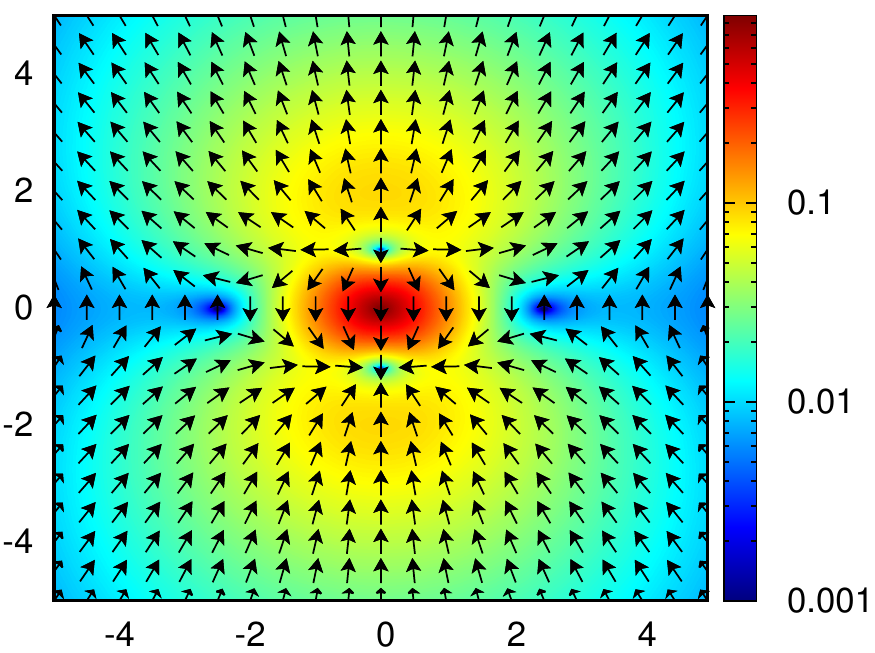,width=7.5cm}
}
\caption{Magnetic vector potential $\bm A \,\frac{4 \pi \ell^2}{\mu_0 m}$, (a) and (c), or $\bm A \,\frac{4 \pi a_1^2}{\mu_0 m}$, (e),  and magnetic field strength $\bm B\,\frac{4 \pi \ell^3}{\mu_0 m}$, (b) and (d), or $\bm B\,\frac{4 \pi a_1^3}{\mu_0 m}$, (f), in classical, (a) and (b), Bopp-Podolsky, (c) and (d), and second gradient electromagnetostatics, (e) and (f), for a magnetic dipole with $\bm m = m \bm e_3$ aligned along the third coordinate direction, for $x_3=0$. Arrows indicate field direction, the color its absolute value. Note that in (a), (b) and (d) the color scale fails to display the singularities.}
\label{fig:b_dip}
\end{figure}

The magnetic field strength~\eqref{B} of a magnetic dipole is given by the curl 
of the magnetic dipole potential~\eqref{A-md1}
\begin{align}
\label{B-md1}
\bm B&=\mu_0\,\nabla\times\big[\bm m\times \nabla\,  G^{L\Delta}\big]\nonumber\\
&=\mu_0 \big[\bm m\, \Delta- (\bm m\cdot \nabla)\,\nabla\big] G^{L\Delta}
\end{align}
and, using Eq.~\eqref{G-BP-grad2}, it becomes
\begin{align}
\label{B-md2}
\bm B&=\frac{\mu_0}{4\pi}\, 
\Big[4\pi\, \bm m\, G^L(R)
+\frac{3(\bm m\cdot\bm R) \bm R}{R^5}\, f_2(R,a_1,a_2)
-
\frac{\bm m}{R^3}\, f_1(R,a_1,a_2)
\Big]\nonumber \\
&=\frac{\mu_0}{4\pi}\, \bigg[
\frac{3(\bm m\cdot\bm R) \bm R}{R^5}\, f_2(R,a_1,a_2)
+\frac{\bm m}{R^3}\, \big[2f_1(R,a_1,a_2)
-3 f_2(R,a_1,a_2)\big]\bigg]\,.
\end{align}
Using the near fields of $f_1$ and $f_2$, Eqs.~\eqref{f1-ser} and \eqref{f2-ser},
it can be seen that the magnetic dipole field~\eqref{B-md2} 
is finite at $R=0$ due to the $f_1$-term, namely  (see Figs.~\ref{fig:f}c and \ref{fig:f}d)  
\begin{align}
\label{B-md2-0}
\bm B(0)=\frac{\mu_0\, \bm m}{6\pi a_1a_2(a_1+a_2)}\,
\end{align}
and it does not have a directional discontinuity.
A comparison of vector potentials and magnetic fields in second gradient electromagnetostatics, the Bopp-Podolsky theory and classical electromagnetostatics 
is given in Fig.~\ref{fig:b_dip}. Again, similarly to the electric dipole fields, the regularization modifies the field direction in the near field.

The magnetic field energy can be written as
\begin{align}
\label{Um}
U_{\text{m}}&=\int_{\Bbb R^3} \frac{1}{\mu_0}
\Big(\bm B \cdot \bm B 
+\ell_1^2 \nabla \bm B :\nabla \bm B 
+\ell_2^4 \nabla\nabla \bm B \mathbin{\vdots} \nabla\nabla \bm B 
\Big)\text{d} V\nonumber\\
&=\int_{\Bbb R^3}   \frac{1}{\mu_0} \, L(\Delta) \bm B \cdot  \bm B\,  \text{d} V\nonumber\\
&=\int_{\Bbb R^3}   \bm H \cdot \bm B\,  \text{d} V\nonumber\\
&=\int_{\Bbb R^3}   \bm J \cdot \bm A\,  \text{d} V\,,
\end{align}
where we have used integration by parts and the surface terms vanish at infinity.
Substituting Eq.~\eqref{md} into Eq.~\eqref{Um},
the interaction energy of two magnetic dipoles $\bm m$ and $\bm m'$, 
a distance $\bm R=\bm r-\bm r'$ apart, 
is given by
\begin{align}
\label{Umm}
U_{\bm m \bm m '}&=\bm m\cdot \bm B_{\bm m'}\nonumber\\
&=\frac{\mu_0}{4\pi}\, \bigg[
\frac{3(\bm m\cdot\bm R) (\bm m' \cdot\bm R)}{R^5}\, f_2(R,a_1,a_2)
+\frac{\bm m\cdot \bm m'}{R^3}\, \big[2f_1(R,a_1,a_2)
-3 f_2(R,a_1,a_2)\big]
\bigg]\,.
\end{align}
The magnetostatic self-energy of a magnetic dipole reduces to
\begin{align}
\label{Um-s}
U_{\text{self}}&=\frac{1}{2}\,U_{\bm m \bm m }(0)\nonumber\\
&=\frac{\mu_0\, (\bm m\cdot\bm m)}{12\pi  a_1a_2(a_1+a_2)}\, ,
\end{align}
which is finite.

The magnetostatic part of the Lorentz force reads as~\citep{Lazar20}
\begin{align}
\label{F-L-m}
\bm{{F}}=\int_{\Bbb R^3} \bm J\times\bm B \, \d V\,.
\end{align}
Substituting Eqs.~\eqref{md} and \eqref{B-md1} into Eq.~\eqref{F-L-m} and using Eqs.~\eqref{G-BP-grad3} and \eqref{G-BP-Lapl-grad},
the magnetostatic interaction force between two magnetic dipoles becomes
\begin{align}
\label{Fmm}
{\bm { F}}_{\bm m \bm m '}
&=\nabla (\bm m\cdot \bm B_{m'})\nonumber\\
&=\mu_0\, \big[(\bm m\cdot\bm m') \nabla\Delta-
(\bm m\cdot \nabla) (\bm m'\cdot \nabla)  \nabla\big]  G^{L\Delta}\nonumber\\
&=\frac{\mu_0}{4\pi}\,
\bigg[\frac{3(\bm m\cdot \bm R) \bm m'+3(\bm m'\cdot \bm R) \bm m + 3(\bm m\cdot \bm m') \bm R}{R^5}\, f_2(R,a_1,a_2)
\nonumber\\
&\ 
-\frac{15(\bm m\cdot\bm R)(\bm m'\cdot\bm R) \bm R}{R^7}\, f_3(R,a_1,a_2)
-\frac{15(\bm m\cdot \bm m') \bm R}{R^5}\, \big[ f_2(R,a_1,a_2)-f_3(R,a_1,a_2)\big]
\bigg]\,,
\end{align}
which is finite at $R=0$ and nonsingular (see Figs.~\ref{fig:f}e and \ref{fig:f}f), 
but it possesses a directional discontinuity at $R=0$.

\subsection{Some qualitative side-effects of regularization}
As we have seen, the gradient theory for large distances approaches the classical results asymptotically, while on small length scales Green functions and therefore electric and magnetic fields are modified. These modifications introduce new qualitative behavior that the reader should be aware of.

For example, apart from weakening or removal of singularities, the field of the electric dipole is reversed in the direct vicinity of the dipole and two singular points appear, zeros of the electric field, where a point charge would be in equilibrium (see Fig.~\ref{fig:e_dip}f). These singular points also exist in the fields of real dipoles, i.e., two point charges a certain distance apart, and thus motivate the following.

Consider three point charges on a straight line, for simplicity we take two charges $q$ at the same distance $r$ to a charge $q'$ in the middle. Due to the symmetry of the situation the force on the charge $q'$ is zero, and through Eq.~\eqref{Fqq} the force on the other two charges is
\begin{align}
\label{Fqq'q}
 F = \frac{qq'}{4\pi\varepsilon_0}\, \frac{1}{r^2}\, \bigg( f_1(r,a_1,a_2) + \frac{q}{4 q'} f_1(2r,a_1,a_2) \bigg)\,.
\end{align}
For $\frac{q}{q'} > 0$ this expression is positive, the force is therefore repulsive for all $r$, while for $\frac{q}{q'} < 0$ we find 
\begin{align}
\label{Fqq'q-large-r}
\bigg( f_1(r,a_1,a_2) + \frac{q}{4 q'} f_1(2r,a_1,a_2) \bigg) \sim 1 + \frac{q}{4 q'}
\end{align}
asymptotically for large $r$ and
\begin{align}
\label{Fqq'q-small-r}
\bigg( f_1(r,a_1,a_2) + \frac{q}{4 q'} f_1(2r,a_1,a_2) \bigg) = \frac{1}{3 a_1a_2 (a_1+a_2)}\, r^3\, \bigg( 1 + \frac{2q}{q'} \bigg) + \mathcal{O}(r^4)
\end{align}
for small $r$. Now, for a charge ratio with $\frac{1}{2} |q'| < |q| < 4 |q'|$, we can have $F$ negative at large distances and positive at sufficiently small distances, by continuity we thus find a zero. This means there exists a situation where three point charges at rest can be in equilibrium just through the electromagnetic forces in vacuum in second gradient electromagnetostatics. This situation has no analogue in classical or Bopp-Podolsky electromagnetostatics.
A linear stability analysis shows that the state is stable with respect to small perturbations in $r$.

\subsection{Reinterpretation of second gradient electromagnetostatics as electromagnetostatics with extended charge and current densities}
Now, we demonstrate that the solutions, Eqs.~\eqref{phi-rp} and \eqref{A-rp}, in  second gradient electromagnetostatics with point and dipole charge and current densities~\eqref{epc},  \eqref{ed} and \eqref{md}
correspond to solutions in classical electromagnetostatics with finite, extended charge and current densities.
Substituting Eq.~\eqref{GBP-conv} into Eqs.~\eqref{phi-rp} and \eqref{A-rp}, we obtain
\begin{align}
\label{phi-rp-L}
\phi&=-\frac{1}{\varepsilon_0}\, G^{\Delta}*\rho^L\,,\\
\label{A-rp-L}
\bm A&=-\mu_0\, G^{\Delta}*\bm J^L\,,
\end{align}
where we have introduced the following extended charge and current densities
\begin{align}
\label{rho-L}
\rho^L&=G^{L}*\rho\,,\\
\label{J-L}
\bm J^L&=G^{L}*\bm J\,,
\end{align}
as convolution of the Green  function $G^{L}$ and the singular, classical charge and current densities $\rho$ and $\bm J$. 
Because the Green function~$G^{L}$ of the bi-Helmholtz operator is finite, also the extended charge and current densities are finite unlike the Bopp-Podolsky case with singular Green function  $G^\text{H}$ of the Helmholtz operator (see Fig.~\ref{fig:GF}). 
The interpretation of extended, singular charge distribution was proposed in the Bopp-Podolsky theory in \citep{Kvasnica,Ji}. 
From the physical point of view, 
the Green function $G^{L}$, Eq.~\eqref{G-BKG-3d}, plays the role of a ``form factor" in Eqs.~\eqref{rho-L} and \eqref{J-L} since it characterizes the shape of the charge and the current. 
So, the electromagnetostatic potentials~\eqref{phi-rp-L} and \eqref{A-rp-L} satisfy the following Poisson equations  
\begin{align}
\label{phi-w-L}
\Delta\,\phi&=-\frac{1}{\varepsilon_0}\, \rho^L\,,\\
\label{A-w_}
\Delta\,\bm A&=- \mu_0\, \BJ^L\,,
\end{align}
with the extended charge and current densities~\eqref{rho-L} and \eqref{J-L} as sources. 
Moreover,
the convolution of the Euler-Lagrange equations~\eqref{EL-1} and \eqref{EL-2}
with the Green function $G^{L}$ and the use of Eqs.~\eqref{KGE},  \eqref{rho-L} and \eqref{J-L} lead to ``classical" inhomogeneous Maxwell equations
\begin{align}
\label{EL-1-L}
&\nabla\cdot \bm E=\frac{1}{\varepsilon_0}\,\rho^L\,,\\
\label{EL-2-L}
&
 \nabla\times\bm B=\mu_0\,\BJ^L\,,
\end{align}
with the extended charge and current densities~\eqref{rho-L} and \eqref{J-L} as sources. 
Of course,  the extended charge and current densities~\eqref{rho-L} and \eqref{J-L} satisfy 
inhomogeneous bi-Helmholtz equations
\begin{align}
\label{rho-L-pde}
L(\Delta)\, \rho^L&=\rho\,,\\
\label{J-L-pde}
L(\Delta)\,\bm J^L&=\bm J\,.
\end{align}
This proves that Eqs. ~\eqref{EL-1-L} and \eqref{EL-2-L} are equivalent to Eqs.~\eqref{EL-1} and \eqref{EL-2}.

\section{Electromagnetic fields
in first gradient electromagnetostatics (Bopp-Podolsky electromagnetostatics)}
\label{sec5}

\begin{figure}[t]\unitlength1cm
\vspace*{0.1cm}
\centerline{
\epsfig{figure=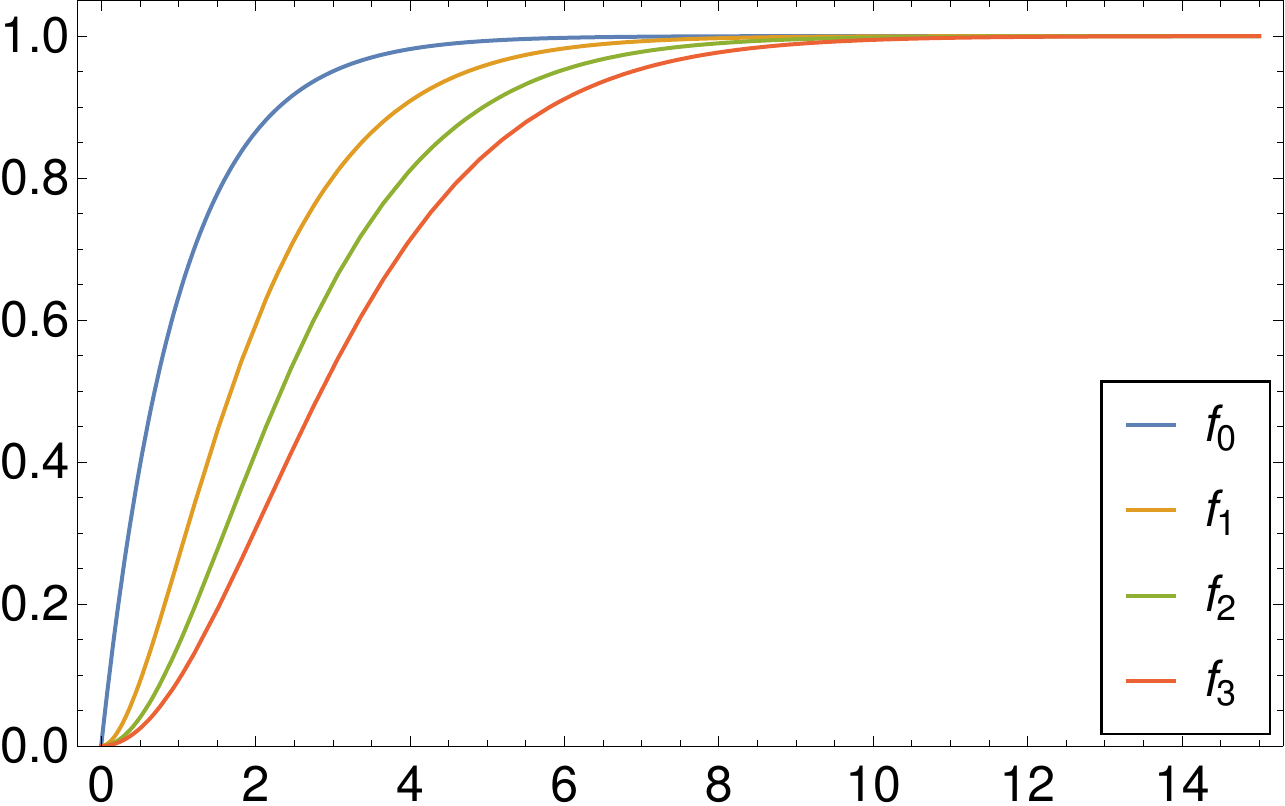,width=7.8cm}
\put(-4.0,-0.4){$R/\ell$}
}
\caption{Plot of the auxiliary functions $f_0$, $f_1$, $f_2$ and $f_3$ in the
  Bopp-Podolsky electromagnetostatics.}
\label{fig:f0-3-BP}
\end{figure}

Now we perform the limit from second gradient electromagnetostatics
to the Bopp-Podolsky electromagnetostatics for the electromagnetic fields of
a point charge, an electric dipole and a magnetic dipole.
In the limit $a_2\rightarrow 0$ and $a_1\rightarrow\ell$ (Bopp-Podolsky limit), 
the auxiliary functions~\eqref{f0}, (\ref{f1})--(\ref{f3}) simplify to
\begin{align}
\label{f0-H} 
f_0(R,\ell)&=1-\e^{-R/\ell}\,,\\
\label{f1-H}
f_1(R,\ell)&=1-\bigg[1+\frac{R}{\ell}\bigg]\,\e^{-R/\ell}\,,
\\
\label{f2-H}
f_2(R,\ell)&=1-
\bigg[1+\frac{R}{\ell}+\frac{1}{3}\,\frac{R^2}{\ell^2}\bigg]\,\e^{-R/\ell}\,,
\\
\label{f3-H}
f_3(R,\ell)&=1-
\bigg[1+\frac{R}{\ell}+\frac{2}{5}\,\frac{R^2}{\ell^2}+\frac{1}{15}\,\frac{R^3}{\ell^3}\bigg]\,\e^{-R/\ell}\,.
\end{align} 
The auxiliary functions~(\ref{f0-H})--(\ref{f3-H})
are plotted in Fig.~\ref{fig:f0-3-BP}.
The relevant series expansion (near field behavior) of the auxiliary
functions~\eqref{f1-H}--\eqref{f3-H} (near fields) reads 
\begin{align}
\label{f0-H-ser}
f_0(R,\ell)&=\frac{1}{\ell}\, R-\frac{1}{2 \ell^2}\, R^2
+\mathcal{O}(R^3)\,,\\
\label{f1-H-ser}
f_1(R,\ell)&=\frac{1}{2\ell^2}\, R^2-\frac{1}{3 \ell^3}\, R^3+\frac{1}{8 \ell^4}\, R^4+\mathcal{O}(R^5)\,,\\
\label{f2-H-ser}
f_2(R,\ell)&=\frac{1}{6\ell^2}\,R^2-\frac{1}{24 \ell^4}\, R^4+\mathcal{O}(R^5)\,,\\
\label{f3-H-ser}
f_3(R,\ell)&=\frac{1}{10\ell^2}\,R^2-\frac{1}{120\ell^4}\, R^4+\mathcal{O}(R^6)\,.
\end{align}
In Eq.~\eqref{f0-H-ser},
it can be seen that the function $f_0(R,\ell)$ regularizes up to a
$1/R$-singularity and gives a nonsingular field expression.
In Eqs.~\eqref{f1-H-ser}--\eqref{f3-H-ser},
it can be seen that the functions $f_1(R,\ell)$,  $f_2(R,\ell)$
and  $f_3(R,\ell)$ regularize up to a $1/R^2$-singularity and give
nonsingular fields. 
At $R=0$ the  auxiliary functions~(\ref{f0-H})--(\ref{f3-H}) are zero and 
in the far field they approach the value of 1.

\subsection{Electric point charge}

In the Bopp-Podolsky limit, the electrostatic potential~\eqref{phi-epc2}
reduces to
\begin{align}
\label{phi-epc2-BP}
\phi=\frac{q}{4\pi\varepsilon_0}\, \frac{1}{R}\, f_0(R,\ell)\,,
\end{align}
which is finite at $R=0$, namely  (see Fig.~\ref{fig:f}a)
\begin{align} 
\label{phi-epc2-BP-0}
\phi(0)=\frac{q}{4\pi\varepsilon_0\,\ell}\,. 
\end{align}
Of course, Eq.~\eqref{phi-epc2-BP} is in agreement with the original expression
given by~\citet{Bopp} and \citet{Podolsky}. 

Moreover, the electric field strength~\eqref{E-epc2}
reduces to
\begin{align}
\label{E-epc2-BP}
\bm E=\frac{q}{4\pi\varepsilon_0}\, \frac{\bm R}{R^3}\, f_1(R,\ell)\,,
\end{align}
which  is in agreement with the expression
given by~\citet{Bopp} (see also \citep{Lande3}). 
The electric field strength of a point charge is nonsingular but has a directional discontinuity 
at the origin, namely
\begin{align}
\label{E-epc2-BP-0}
\bm E(\bm R)=\frac{q}{8\pi\varepsilon_0 \ell^2}\, \hat{\bm R}\, + \mathcal{O}(R)\,,
\end{align}
where $ \hat{\bm R}=\bm R/R$. 
Therefore, the projection of the electric field~\eqref{E-epc2-BP} onto a curve passing trough the location of the charge jumps
from $q/(8\pi\varepsilon_0 \ell^2)$ to 
$-q/(8\pi\varepsilon_0 \ell^2)$ at $\bm R=0$. 

The self-energy~\eqref{Uq} of a point charge becomes
\begin{align}
\label{Uq-BP}
U_{\text{self}}=\frac{q^2}{8\pi\varepsilon_0\, \ell}\,,
\end{align}
which is positive and finite for $\ell>0$.

\subsection{Electric dipole}

In the Bopp-Podolsky limit, the electrostatic potential~\eqref{phi-ed2}
reduces to
\begin{align}
\label{phi-ed2-BP}
\phi=\frac{1}{4\pi\varepsilon_0}\, \frac{\bm p\cdot \bm R}{R^3}\, f_1(R,\ell)\,,
\end{align}
which is in agreement with the expression given by~\citet{Bonin2019}.
The electrostatic potential of an electric dipole is finite and possesses a directional discontinuity 
at the origin, namely
\begin{align}
\label{phi-ed2-BP-0}
\phi(\bm R)=\frac{1}{8\pi\varepsilon_0 \ell^2}\, (\bm p \cdot \hat{\bm R}) + \mathcal{O}(R)\,.
\end{align}

The electrostatic field strength~\eqref{E-ed2} becomes
\begin{align}
\label{E-ed2-BP}
\bm E=\frac{1}{4\pi\varepsilon_0}\, 
\bigg[
\frac{3(\bm p\cdot\bm R) \bm R}{R^5}\, f_2(R,\ell)
-\frac{\bm p}{R^3}\, f_1(R,\ell)
\bigg]\,,
\end{align}
which is singular because it has a $1/R$-singularity at the origin.
Therefore, the self-energy~\eqref{Up} of an electric dipole and the interaction force~\eqref{Fpp} between two electric 
dipoles become infinite and singular in the Bopp-Podolsky theory.

\subsection{Magnetic dipole}

In the Bopp-Podolsky limit, the magnetic vector potential~\eqref{A-md2}
reduces to
\begin{align}
\label{A-md2-BP}
\bm A=\frac{\mu_0}{4\pi}\, \frac{\bm m\times \bm R}{R^3}\, f_1(R,\ell)\,,
\end{align}
which is in agreement with the expression given by~\citet{Bonin2019}.
The magnetostatic potential of a magnetic dipole is 
finite and possesses a directional discontinuity 
at the origin 
\begin{align}
\label{A-ed2-BP-0}
\bm A(\bm R)=\frac{\mu_0}{8\pi \ell^2}\, (\bm m \times \hat{\bm R}) + \mathcal{O}(R)\,.
\end{align}

The magnetic field strength~\eqref{B-md2} becomes
\begin{align}
\label{B-md2-BP}
\bm B&=\frac{\mu_0}{4\pi}\, 
\Big[4\pi\, \bm m\, G^{\text{H}}(R)
+\frac{3(\bm m\cdot\bm R) \bm R}{R^5}\, f_2(R,\ell)
-
\frac{\bm m}{R^3}\, f_1(R,\ell)
\Big]\nonumber\\
&=\frac{\mu_0}{4\pi}\, \bigg[
\frac{3(\bm m\cdot\bm R) \bm R}{R^5}\, f_2(R,\ell)
+\frac{\bm m}{R^3}\, \big[2f_1(R,\ell)
-3 f_2(R,\ell)\big]
\bigg]\,,
\end{align}
which is singular since it possesses a $1/R$-singularity at the origin.
Therefore, the self-energy~\eqref{Um} of a magnetic dipole 
and the interaction force~\eqref{Fmm} between two magnetic dipoles 
become infinite and singular in the Bopp-Podolsky theory.
Note that Eq.~\eqref{B-md2-BP} is in agreement with the expression given by~\citet{Bonin2019}.

\section{Electromagnetic fields
in classical Maxwell electromagnetostatics}
\label{sec6}

The limit from the Bopp-Podolsky electromagnetostatics to the classical 
Maxwell electromagnetostatics is $\ell\rightarrow 0$. 

\subsection{Electric point charge}

From Eqs.~\eqref{phi-epc2-BP} and \eqref{E-epc2-BP}, 
the classical electrostatic potential and electric field strength
of a point charge
\begin{align}
\label{phi-epc2-M}
\phi=\frac{q}{4\pi\varepsilon_0}\, \frac{1}{R}
\end{align}
and
\begin{align}
\label{E-epc2-M}
\bm E=\frac{q}{4\pi\varepsilon_0}\, \frac{\bm R}{R^3}\,,
\end{align}
respectively, are recovered.

\subsection{Electric dipole}

From Eqs.~\eqref{phi-ed2-BP} and \eqref{E-ed2-BP}, 
the classical electrostatic potential and the electric field strength
of an electric dipole are obtained as
\begin{align}
\label{phi-ed2-M}
\phi=\frac{1}{4\pi\varepsilon_0}\, \frac{\bm p\cdot \bm R}{R^3}
\end{align}
and 
\begin{align}
\label{E-ed2-M}
\bm E=\frac{1}{4\pi\varepsilon_0}\, 
\bigg[
\frac{3(\bm p\cdot\bm R) \bm R}{R^5}
-\frac{\bm p}{R^3}
-\frac{4\pi}{3}\, \bm p\, \delta(\bm R)
\bigg]\,,
\end{align}
respectively. 
Note that Eqs.~\eqref{phi-ed2-M} and \eqref{E-ed2-M} are in agreement with 
the expressions\footnote{ 
In the classical Maxwell electromagnetostatics of dipoles, 
the  term proportional to the Dirac delta function 
follows from the second order derivatives of $1/R$ 
in the sense of generalized functions~\citep{Kanwal}
and is known as the Frahm formula~\citep{Frahm}
\begin{align*}
\nabla\nabla\bigg(\frac{1}{R}\bigg)=
\frac{3\bm R\bm R}{R^5} -\frac{\bm 1}{R^3}
-\frac{4\pi}{3}\, \bm 1\, \delta(\BR)\,.
\end{align*}
}
given in the literature~\citep{Jackson,Frahm,Griffiths,Leung}.

\subsection{Magnetic dipole}

From Eqs.~\eqref{A-md2-BP} and \eqref{B-md2-BP}, 
the classical magnetic vector potential and the magnetic field strength
of a magnetic dipole are obtained as
\begin{align}
\label{A-md2-M}
\bm A=\frac{\mu_0}{4\pi}\, \frac{\bm m\times \bm R}{R^3}
\end{align}
and
\begin{align}
\label{B-md2-M}
\bm B&=\frac{\mu_0}{4\pi}\, 
\bigg[
\frac{3(\bm m\cdot\bm R) \bm R}{R^5}
-\frac{\bm m}{R^3} 
+\frac{8\pi}{3}\, \bm m\, \delta(\bm R)
\bigg]\,,
\end{align}
respectively. 
Note that Eqs.~\eqref{A-md2-M} and \eqref{B-md2-M} are in agreement with 
the expressions given in the literature~\citep{Jackson,Frahm,Griffiths,Leung}.

\section{Conclusion}
\label{concl}

We have presented second gradient electromagnetostatics,
which is the static version of second gradient electrodynamics, a generalization of Bopp-Podolsky electrodynamics, 
as singularity-free field theory of electromagnetic fields, in particular, for an electrostatic point charge, an electrostatic dipole 
and a magnetostatic dipole. 
Through linear field equations of sixth order, second gradient electromagnetostatics yields an even stronger regularization than the Bopp-Podolsky theory: the field of a point charge is nonsingular and zero at the origin, the fields of electric and magnetic dipoles are nonsingular as well and dipoles have a finite self-energy, which diverges in the limit to Bopp-Podolsky electrodynamics (see Table~\ref{table}).

The regularization stems from the fact that the Green function of the bi-Helmholtz-Laplace operator appearing in second gradient  electromagnetostatics
and its first, second, and third gradients are singularity-free:
 \begin{align}
\label{reg}
G^{L\Delta}&=\text{reg}\, \big[G^\Delta\big]\,,\\
\label{reg-grad}
\nabla G^{L\Delta}&=\text{reg}\, \big[\nabla G^\Delta\big]\,,\\
\label{reg-grad2}
\nabla\nabla G^{L\Delta}&=\text{reg}\, \big[\nabla\nabla G^\Delta\big]\,,\\
\label{reg-grad3}
\nabla\nabla\nabla G^{L\Delta}&=\text{reg}\, \big[\nabla\nabla\nabla G^\Delta\big]\,.
\end{align}
In every single problem studied, we were able to recover the Bopp-Podolsky results and the Maxwell results in the proper limits, 
when the length scale parameter  $\ell_2$ (or $a_2$) goes to zero and the Bopp-Podolsky length scale parameter $\ell$ goes to zero, respectively. 

\begin{table}[t]
\caption{Comparison of the near-field behavior of the 
electromagnetic fields of an electric point charge, an electrostatic dipole and a magnetostatic dipole.}
\begin{center}
\leavevmode
\begin{tabular}{||c|c|c|c|c||}\hline
Theory & \multicolumn{2}{|c|}{electric point charge}  
& 
\multicolumn{2}{|c||}{electric and magnetic dipoles}\\
\hline
& $\phi$& $\bm E$&  $\phi$, $\bm A$ & $\bm E$, $\bm B$\\
\hline
Maxwell theory& $1/R$&  $1/R^2$ &   $1/R^2$& $1/R^3$ and $\delta(\bm R)$\\
Bopp-Podolsky theory & finite & discontinuity &  discontinuity & $1/R$ \\
Second gradient theory & finite & approaching zero & approaching zero& finite\\
\hline
\end{tabular}
\end{center}
\label{table}
\end{table}

We have also demonstrated that, much like in Bopp-Podolsky electrodynamics, the above results can be obtained through a special ansatz for extended charge distributions in classical electrostatics instead of an invariant generalized field theory. This is due to the structure of the Green function, a convolution of multiple Green functions to second-order operators, some of which can be attributed to a charge distribution in this interpretation. 

Another interesting detail in second gradient electromagnetostatics is the existence of stable equilibria of three point charges. While in classical electrodynamics the maximum principle for the Poisson equation forbids such states (this is sometimes referred to as Earnshaw's theorem), the higher-order generalized analogue to the Poisson equation has no maximum principle. The same is true for the fourth-order equation for the potential in Bopp-Podolsky electrostatics, although a stable configuration of point charges might be harder to construct there.

Finally, we conclude that the theory of second gradient electromagnetostatics provides a singularity-free generalized continuum theory of 
electromagnetostatics with generalized Gauss law and generalized Amp{\`e}re law valid down to short distances.
The covariant form of second gradient electrodynamics and its meaning as a nonsingular relativistic field theory will be given in  a forthcoming publication.

\section*{Acknowledgement}
M.L. gratefully acknowledges the grant from the 
Deutsche Forschungsgemeinschaft (Grant No. La1974/4-1). 
J.L. wishes to express his gratitude to Markus Lazar for supervision of his master's thesis and for proposing the subject.

\end{document}